\documentclass[10pt,twocolumn,twoside,floatfix]{revtex4}
\usepackage[T1]{fontenc}
\usepackage[utf8]{inputenc}
\usepackage{textcomp}
\usepackage{fourier}
\usepackage{graphicx}
\usepackage[usenames,dvipsnames]{xcolor}
\usepackage{xr-hyper}
\usepackage{hyperref}
\usepackage[final]{showlabels}

\newcommand{\tens}[1]{\bar{\bar{#1}}}

\newcommand{\irm}{\mathrm{i}}
\newcommand{\erm}{\mathrm{e}}
\newcommand{\drm}{\mathrm{d}}
\newcommand{\TM}{\mathrm{\parallel}}
\newcommand{\TE}{\mathrm{\perp}}
\newcommand{\alphaTM}{\alpha^{\TM}}
\newcommand{\alphaTE}{\alpha^{\TE}}
\newcommand{\phiTM}[1][]{\varphi^{#1\TM}}
\newcommand{\psiTM}[1][]{\psi^{#1\TM}}
\newcommand{\phiTE}[1][]{\varphi^{#1\TE}}
\newcommand{\psiTE}[1][]{\psi^{#1\TE}}
\newcommand{\matr}[4]{\left(\begin{array}{cc}#1&#2\\#3&#4\end{array}\right)}

\newcommand{\MV}{\mathcal{V}}
\newcommand{\ML}{\mathcal{L}}
\newcommand{\Mint}{\mathcal{I}}
\newcommand{\Mpro}{\mathcal{P}}
\newcommand{\bigO}[1]{\mathcal{O}\left(#1\right)}
\newcommand{\chem}[1]{\(\mathrm{#1}\)}

\newcommand{\suprefsec}[1]{Suppl. Inf.~\cite[Sec.~\ref{S:#1}]{SupplementalMaterial}}

\newcommand{\AIM}{Anisotropic Interface model}
\newcommand{\ALM}{Anisotropic Layer model}
\newcommand{\MDM}{Microscopic dipolar model}
\newcommand{\ITM}{Isotropic thin-film model}
\renewcommand{\AIM}{AIM}
\renewcommand{\ALM}{ALM}
\renewcommand{\MDM}{MDM}
\renewcommand{\ITM}{ITM}

\renewcommand{\AIM}{AnisInt model}
\renewcommand{\ALM}{AnisLay model}
\renewcommand{\MDM}{MicroDip model}
\renewcommand{\ITM}{IsoFilm model}

\graphicspath{{Figs-pdf/}}

\providecommand{\affiliation}[1]{}
\begin{document}

\title{Optical modelling of single and multilayer 2D materials and heterostructures}

\author{Bruno Majérus}
\affiliation
{Department of Physics \& Namur Institute of Structured Matters (NISM), University of Namur, 61 rue de Bruxelles, B-5000 Namur, Belgium.}
\author{Luc Henrard}
\affiliation
{Department of Physics \& Namur Institute of Structured Matters (NISM), University of Namur, 61 rue de Bruxelles, B-5000 Namur, Belgium.}
\author{Pascal Kockaert}
\affiliation
{OPERA-photonics, Université libre de Bruxelles (U.L.B.), 50 Avenue F. D. Roosevelt, CP 194/5, B-1050 Bruxelles, Belgium}
\email{Pascal.Kockaert@ulb.ac.be}

\begin{abstract}
Bidimensional materials are ideally viewed as having no thickness, as their name suggests. Their optical response have been previously modelled by a purely bidimensional surface current or by a very thin film with some contradictory results. The advent of multilayer stacks of bidimensional materials and combinations of different materials in vertical van der Waals heterostructures highlights however that these materials have a finite thickness. In this article, we propose a new model that reconciles both approaches and we show how volume properties of stacked bidimensional layers can be calculated from the bidimensional response of each individual layer, and conversely. In our approach, each bidimensionnal layers is surrounded by vacuum and described as a kind of transfer matrix with intrinsic parameters that do not depend on the external medium. This provides a link between continuous thin films and discrete layers. We show how to model heterostructures of bidimensional materials and identify the parameters of the current sheet that represents the bidimensional material in the zero-thickness limit, namely the in-plane surface susceptibility and the out-of-plane displacement susceptibility.

We show that our unified model is perfectly compatible with existing ellipsometric data with the same reliability as the existing interface model but with different values of the surface susceptibility or bulk dielectric function. We discuss in details the origin of the discrepancies and show that our approach allows to determine intrinsic properties of the bidimensional materials with the advantage that multilayer and monolayer systems are described in a same framework.
\end{abstract}

\maketitle

\section{Introduction}
Following the important impact of 2D materials in different domains of applications such as electronics, photonics, biomedical engineering, printing technology and aerospace~\cite{dhinakaran_review_2020}, multilayer systems are progressively developing. They include vertical van der Waals heterostructures~\cite{geim_van_2013,novoselov_2d_2016} and could be used to increase the total layer response~\cite{song_complex_2019,wei_layer-dependent_2022,fang_layer-dependent_2020,fang_thickness_2022}, study the interaction between layers~\cite{song_complex_2019}, protect against electromagnetic microwave pollution~\cite{bai_mass_2019}, or to devise new materials combining their properties by stacking them together.

Optical properties of single- and multilayer systems can be used to probe the number of layers~\cite{graf_spatially_2007,zhang_optical_2021} or to investigate the evolution of the refractive and conductive properties with the number of layers~\cite{song_complex_2019,wei_layer-dependent_2022,fang_layer-dependent_2020,fang_thickness_2022,Majerus_18}. In waveguiding optics, 2D materials or multilayer systems are useful for functionalization~\cite{chang_graphene-integrated_2022}, modulation~\cite{sun_optical_2016,gan_2d_2022} or to enhance the nonlinear properties~\cite{alexander_electrically_2017,vermeulen_graphenes_2018,dremetsika_enhanced_2017,liu_measuring_2018}.

\begin{figure}
\includegraphics[width=\linewidth]{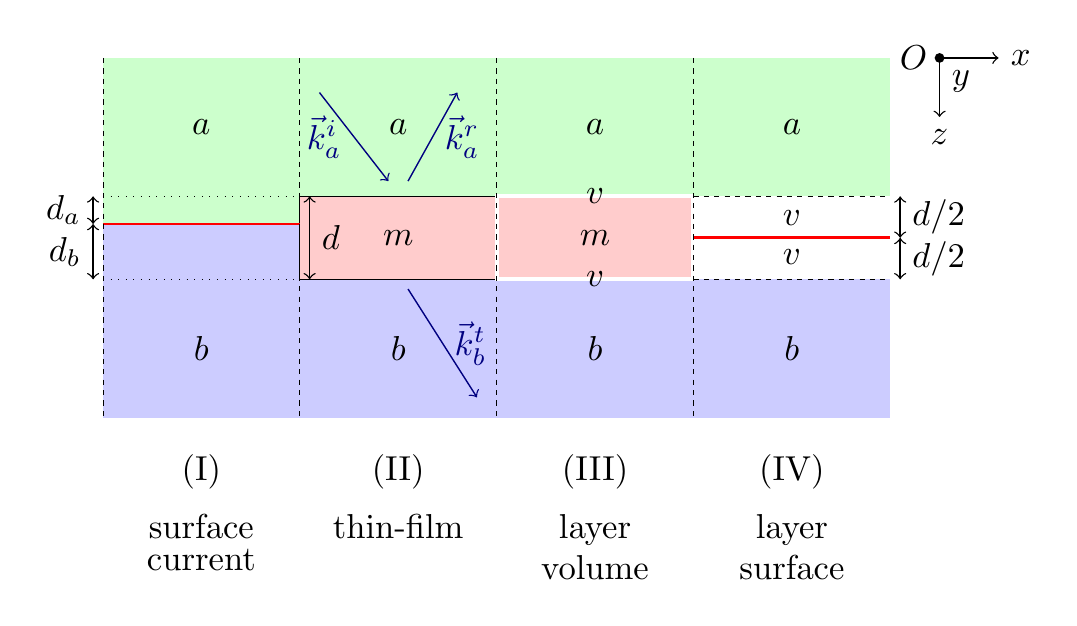}
\caption{Geometries considered in different models.
(I) surface current used in {\AIM} and {\MDM};
(II) thin-film used in {\ITM} ;
(III)-(IV) continuous and discrete layer geometries used in our new {\ALM}, and compatible with multi-layer systems.  Surrounding materials are labeled \(a\) and \(b\).
The interface (I) is described with a surface susceptibility \(\chi^s\) (red line).
The thin-film (II) is described by a permittivity \(\varepsilon_m\) and a thickness \(d\).
The discrete layer (III\,\&\,IV) is modeled using either a bulk permittiviy \(\varepsilon_m\) or a current sheet with surface susceptibility \(\chi^s\) (red line). Each layer is surrounded with vacuum (\(v\)) and has a total thickness \(d\).}
\label{Fig:sc-tf-lay}
\end{figure}

With their atomic thickness, 2D materials can be described either by surface currents at an interface or as ultra-thin films of continuous materials. This is depicted on  Fig.~\ref{Fig:sc-tf-lay} that presents three approaches: (I) the surface current or interface; (II) the thin film; and (III)-(IV) our new anisotropic layer model in continuous and discrete versions. Up to now, the surface current and thin-film descriptions have been used in parallel, raising questions about which model is more physically sound and why they differ on their predictions~\cite{matthes_influence_2016}.

A first attempt to merge the two models in a single picture was provided by Majérus \textit{et al.} in~\cite{majerus_electrodynamics_2018} within an Anisotropic Interface Model (\AIM), where the importance of the anisotropy of the thin-film model and of the out-of-plane component of the surface susceptibility was pointed out. In a very nice experimental work~\cite{xu_optical_2021}, Xu and co-workers highlighted the importance of the out-of-plane component to interpret the measurements. More recently, Dell'Anna showed how to model this out-of-plane response from a microscopic dipolar point of view in a Microscopic Dipolar Model (\MDM)~\cite{dellanna_reflection_2022}. Although the works in \cite{majerus_electrodynamics_2018} and~\cite{xu_optical_2021} present similarities, they do not provide the same analytical expressions for the reflection coefficient, and the first of these models is not purely conservative for a lossless 2D material immersed in a lossless medium as pointed out by~\cite{xu_optical_2021}. Another major drawback of both approaches is that they do not provide intrinsic electromagnetic response quantities, independently of the surrounding medium.  More over, while these two works describe the impact of a 2D material at the \emph{interface} between two continuous bulk media, they do not provide a clear way to extend the models to \emph{multilayer} systems. Tab.~\ref{Tab:sc-tf-lay} summarizes the main characteristics of the different models that are depicted on Fig.~\ref{Fig:sc-tf-lay}. It shows that existing models, such as the {\AIM} and {\MDM}  describe well the boundary conditions at an interface, but cannot be extended to the multilayer systems; while the Isotropic Thin-Film model (\ITM) can describe multilayer systems but without anisotropy.

\begin{table}
\caption{
Ability of the main models to describe:\\
(i) an \emph{interface} with zero-thickness;\\
(ii) a \emph{single layer} with a non-zero thickness;\\
(iii) \emph{multilayer} stacks of 2D materials;\\
(iv) the in-plane -- out-of-plane \emph{anisotropy} of the material;\\
(v) \emph{flux conservation} in lossless systems.\\
Only our new AnisLay model combines all features.}
\label{Tab:sc-tf-lay}

\strut

\footnotesize

\begin{tabular}{r|cccc}
Model & AnisInt & MicroDip & IsoFilm & AnisLay \\
\hline
interface    & + & + & - & + \\
single layer & - & - & + & + \\
multi-layer  & - & - & + & + \\
anisotropy   & + & + & - & + \\
flux conservation & - & + & + & +
\end{tabular}

\strut

\footnotesize

\begin{tabular}{r|l|l}
Model &\multicolumn{1}{|c|}{Full name}& Ref\\
\hline
AnisInt & Anisotropic Interface &\cite{majerus_electrodynamics_2018};\\
MicroDip & Microscopic Dipolar &\cite{dellanna_reflection_2022,xu_optical_2021};\\
IsoFilm & Isotropic Thin-Film &\cite{Li_Heinz_18};\\
AnisLay & Anisotropic Layer&(Sec.~\ref{Sec:layermodel}).
\end{tabular}

\end{table}

In this work, we propose a new model that combines all the features mentioned above. Heterostructures made of successive stacking of several 2D layers are also described in the approach. These expressions will be cast on the convenient form of a transfer matrix that is very useful for theoretical descriptions or for applications like ellipsometry. In doing this, we will also show how the {\AIM} of Majérus~\cite{majerus_electrodynamics_2018} and the {\MDM} of Dell'Anna~\cite{dellanna_reflection_2022} differ. Our model also identifies \textit{intrinsic parameters} of a layer, \textit{i.e.} parameters that do not depend on the surrounding media. This condition is mandatory for a general model but it was not encountered previously.

Our approach will be analytical in the first place and based on the description of a single sheet of 2D material: the Anisotropic Layer Model ({\ALM}). It will be compared with existing experimental results. We will show that the different proposed models can fit the experimental data quite well with, surprisingly, different conclusions on the deduced response functions (susceptibility and dielectric tensor). Our model is shown to be more general, as it is the only one that fits experimental results and describes a N-layer material as a (N-1)-layer material plus a single layer.

Our analysis will be performed assuming that there is no change of the intrinsic parameters with the number of layers or the heterostructure stacking. We will on purpose ignore inter-layer effects as the charge transfer and the covalent bounding ~\cite{song_complex_2019} that could lead to new physics related to the stacking order or orientation. This will allow to identify such interactions through the comparison of  experimental data on multilayer systems with our optical model. Such effects of the interaction between bidimensional materials are however restricted close to the Fermi level and do not influence the electromagnetic properties in the near infrared and the visible.

In Section~\ref{Sec:ContCond}, we introduce the existing interface 2D layer models. Then, in Sec.~\ref{Sec:layermodel}, we present our new model, defined for a single anisotropic layer of 2D material and applied thereafter to an homogeneous multilayer system. After that, we relate the new model to the existing ones, showing how the new layer model leads to the existing interface model in the zero-thickness limit. In Sec.~\ref{Sec:hybridmat}, we apply the new model to build the response of a simple heterostructure and identify the parameters that should be used to describe 2D materials independently of the surrounding media. Then in Sec.~\ref{Sec:fitting}, we show that our model is compatible with existing data on single-layer graphene and \chem{MoS_2}, with the same reliability as the previous models, but with the advantage that it makes use of intrinsic parameters. We extend our analysis to stacked 2D-layers in 10-layer and 100-layer structures of graphene and \chem{MoS_2}, and we highlight the importance of the choice of geometrical arrangement of the different layers to model as precisely as possible the real distribution of dipolar response in the structure. Finally, we summarize our main findings and conclude.

\section{The existing interface 2D layer models}
\label{Sec:ContCond}
\label{Sec:interfacematrix}

Continuity conditions and translational invariance allow to calculate the evolution of an electromagnetic wave at a planar interface between homogeneous media. In particular, without surface currents,  tangential components of the electric and magnetic fields are continuous, as well as  normal components of the induction and displacement fields.
This is not valid anymore in the presence of an interface 2D layer. Indeed, a surface polarization field orthogonal to the interface \(\mathcal{P}_{\perp}\) appears in this case, as shown in~\cite{sipe_new_1987,dremetsika_enhanced_2017,majerus_electrodynamics_2018} using a macroscopic model, and in~\cite{xu_optical_2021,dellanna_reflection_2022} using a microscopic dipolar approach involving reaction fields.

This surface polarisation \(\vec{\mathcal{P}}\) is related to the electric field  at the interface \( \vec{\mathcal{P}}\) by the surface polarisability tensor \(\tens{\chi}_{s}\) through  \(\vec{\mathcal{P}} = \varepsilon_0 \tens{\chi}_{s} \vec{E}_s\). In their microscopic approach Dell'Anna \textit{et al.}~\cite{dellanna_reflection_2022} consider the surface field equal to the transmitted electric field while in the continuum approach of ~\cite{majerus_electrodynamics_2018} and ~\cite{felderhof_electromagnetic_1987}, the average field of the incident and transmitted regions was used. This choice then influences the obtained value of \(\chi_s\) that is then not intrinsic to the layer but depends also on the surrounding media.  Note that others authors do not explicitly evaluate this field ~\cite{sipe_new_1987,dremetsika_enhanced_2017}.

Later, in Sec.~\ref{Sec:layermodel}, we consider the continuity of the displacement field by introducing the out-of-plane displacement susceptibility \(\xi_z\), and its surface counterpart
\begin{equation}
\xi_z^s = d\, \xi_z = d\,\frac{P_z}{D_z} =  d\,\frac{\varepsilon_z-\varepsilon_0}{\varepsilon_z},
\label{Eq:xizs}
\end{equation}
with \(d\) the layer thickness, \(P_z\), \(D_z\) and \(\varepsilon_z\) respectively the out-of-plane polarization field, displacement field, and permittivity. This resolves the arbitrary choice made in other approaches.

Here below, we summarize results published in~\cite{majerus_electrodynamics_2018} and compare them to those reported in~\cite{dellanna_reflection_2022}, as we will start from these results to build our new layer model.
The geometry  is depicted on Fig.~\ref{Fig:sc-tf-lay}(I). As detailed in~\cite{majerus_electrodynamics_2018}, and summarized in appendix~\ref{Sec:BCTETM}, anisotropic boundary conditions at a 2-D interface between media \(a\) and~\(b\), on  which a monochromatic plane wave is incident, write
\begin{eqnarray}
 1-\frac{r}{t}-\frac{1}{t}&=&\irm{}\frac{\varphi_{ab}}{t},\label{Eq:BC:x}\\
 \alpha_{ab}+\frac{r}{t}-\frac{1}{t}&=&\irm\frac{\psi_{ab}}{t},\label{Eq:BC:yz}
\end{eqnarray}
where the coefficients of transmission \(t\) and reflection \(r\) and the parameter \(\alpha_{ab}\) are defined differently for transverse-electric (TE, \(\TE\)) and transverse-magnetic (TM, \(\TM\)) waves [see (\ref{Eq:kzTE}) and (\ref{Eq:kzTM})]. To avoid coupling between TE and TM waves, we assume that the permittivity tensor is diagonal in the \(Oxyz\) reference frame (see Fig.~\ref{Fig:sc-tf-lay}). This condition is fulfilled by all bidimensional materials with a symmetry axis orthogonal to the 2D plane of order 3, 4 or 6,  which includes graphene and \chem{MoS_2}.
The complex phase shifts \(\varphi_{ab}\) and \(\psi_{ab}\) are defined in appendix.

A transfer matrix links the forward (\(F\)) and backward (\(B\)) components of a plane wave, so that the components in the input medium (\(a\)) are obtained by multiplying the matrix of the system \(\mathcal{M}_{a\rightarrow{b}}\) with the components at the output medium (\(b\)).
\begin{equation}
\left(\begin{array}{c}F_a\\B_a\end{array}\right)
= \frac{1}{t}\left(\begin{array}{cc}1&-r^{\prime}\\r&tt^{\prime}-rr^{\prime}\end{array}\right)
\left(\begin{array}{c}F_b\\B_b\end{array}\right),
\label{Eq:trmat}
\end{equation}
where \(r\) and \(t\) are the transmission and reflection coefficients appearing in (\ref{Eq:BC:x}) and~(\ref{Eq:BC:yz}), and their primed versions are calculated by swapping indices \(a\) and~\(b\).

Transfer matrix coefficients are easy to calculate if we rewrite (\ref{Eq:BC:x}) and~(\ref{Eq:BC:yz}) as
\begin{eqnarray}
 \frac{1}{t}&=&\frac{1+\alpha_{ab}}{2}
 -\irm\frac{\varphi_{ab}}{2t}
 -\irm\frac{\psi_{ab}}{2t},\label{Eq:1overt}\\
 \frac{r}{t}&=&\frac{1-\alpha_{ab}}{2}
 -\irm\frac{\varphi_{ab}}{2t}
 +\irm\frac{\psi_{ab}}{2t}.\label{Eq:rovert}
\end{eqnarray}

This form is also very interesting to compare analytical predictions of different models, as they are usually provided in terms of \(r\) and \(t\) coefficients.
In particular, when the 2D material is immersed in a single material \(\alpha_{ab}=\alpha_{aa}=1\),
\begin{eqnarray}
 t&=& t^\prime = 1
 +\irm\frac{\varphi_{aa}}{2}
 +\irm\frac{\psi_{aa}}{2},\label{Eq:timmersedO1}\\
r&=&  r^\prime =
 -\irm\frac{\varphi_{aa}}{2}
 +\irm\frac{\psi_{aa}}{2}.\label{Eq:rimmesredO1}
\end{eqnarray}
These expressions can directly be compared to those of the {\MDM}~\cite{xu_optical_2021}. Although these equations differ, they are the same at first order in \(\varphi_{ab}\) and \(\psi_{ab}\), if the surface polarisability is defined from the transmitted field in region \(b\) as \(\vec{\mathcal{P}} = \varepsilon_0\tens{\chi}\vec{E}_{b}\) (see~\suprefsec{Sec:microscopic_vs_lay}).

\section{The new anisotropic layer model for single- and multi-layer systems}
\label{Sec:layermodel}

In what precedes, we have written the transfer matrix for a strictly 2D interface. It cannot be used to describe a bulk material made of stacked 2D layers, as it would result also in a strictly bidimensional material with a total zero thickness. We therefore define an anisotropic layer with finite thickness but still strongly related to the interface model, that will prove an interesting alternative to the description in terms of interfaces and propagation matrices.

This is why we introduce a new way to describe light matter interactions.
We will first define a layer and provide its transfer matrix expression.
We will then use the concept of layer in the frame of a continuous medium and in the frame of a discrete medium, and we will show to which extent these two different visions are compatible, and provide an estimate of the precision required on phase measurements to distinguish between these two approaches.

\subsection{Single layer}
We define a single layer as a system of thickness \(d\) surrounded by vaccuum, and providing a dipolar contribution that can either be described as a continuous volume contribution [Fig.~\ref{Fig:sc-tf-lay}(III)], or as a discrete surface contribution [Fig.~\ref{Fig:sc-tf-lay}(IV)].

We start with the volume contribution characterized by an effective permittivity tensor \(\tens{\varepsilon}\). As in Sec.~\ref{Sec:interfacematrix}, we assume that this tensor is diagonal when written in the axes \(Oxyz\).
This layer includes interface matrices to model a homogeneous anisotropic material surrounded by vacuum.  Denoting by \(\Mint_{mn}\) the usual interface matrix separating homogeneous media \(m\) and \(n\) and by \(\mathcal{P}_m(d)\) the propagation matrix in medium \(m\) over a distance \(d\), the layer matrix is
\begin{eqnarray}
\lefteqn{\ML_m(d)}\nonumber\\
&=&
\Mint_{vm}
\Mpro_{m}(d)
\Mint_{mv}
\label{Eq:LmProd}\\
&=&
\matr{\frac{1+\alpha}{2}}
     {\frac{1-\alpha}{2}}
     {\frac{1-\alpha}{2}}
     {\frac{1+\alpha}{2}}
\matr{\erm^{-\irm\Phi}}
     {0}
     {0}
     {\erm^{\irm\Phi}}
\matr{\frac{\alpha+1}{2\alpha}}
     {\frac{\alpha-1}{2\alpha}}
     {\frac{\alpha-1}{2\alpha}}
     {\frac{\alpha+1}{2\alpha}}
     \label{Eq:LmDetaileProd}\\
&=&
\matr{\cos\Phi-\irm\frac{1+\alpha^2}{2\alpha}\sin\Phi}
     {\irm\frac{1-\alpha^2}          {2\alpha}\sin\Phi}
     {-\irm\frac{1-\alpha^2}         {2\alpha}\sin\Phi}      {\cos\Phi+\irm\frac{1+\alpha^2}{2\alpha}\sin\Phi}\label{Eq:LayerMatrix},
\end{eqnarray}
where we have defined \(\alpha=\alpha_{vm}=1/\alpha_{mv}\) and written \(\Phi = k_{z}^m d\) the phase shift induced by the propagation. Detailed expressions for \(\alpha_{mn}\) and \(k_z^m\) are given in appendix, Eqs.~(\ref{Eq:kzTE}) and (\ref{Eq:kzTM}).

We will show that the matrix of a layer of thickness \(d\), seen as a continuous medium [Fig.~\ref{Fig:sc-tf-lay}(III)] is equivalent to the matrix of a zero-thickness current sheet surrounded by two vacuum layers of thickness \(d/2\) [Fig.~\ref{Fig:sc-tf-lay}(IV)]. This explains the dual representation of the {\ALM} on Fig.~\ref{Fig:sc-tf-lay}.

It is important to notice that usual thin film models [Fig.~\ref{Fig:sc-tf-lay}(II)] differ from our definition of a layer as they consider a transfer matrix \(\mathcal{M}_f=\Mint_{am}
\Mpro_{m}(d)
\Mint_{mb}\). This prevents to combine different thin film matrices easily to form a multilayer by matrix multiplication.

To the contrary the layer matrices can be combined straightforwardly to make a thicker material as follows
\begin{eqnarray}
\lefteqn{\ML_m(d_1)\ML_m(d_2)}\nonumber\\
&=&
{\left[\Mint_{vm}
\Mpro_{m}(d_1)
\Mint_{mv}\right]
\left[\Mint_{vm}
\Mpro_{m}(d_2)
\Mint_{mv}\right]}\nonumber\\
&=&
\Mint_{vm}
\Mpro_{m}(d_1+d_2)
\Mint_{mv}\label{Eq:prodI1I2}\\
&=&\ML_m(d_1+d_2),
\end{eqnarray}
as \(\Mint_{mv} = \Mint_{vm}^{-1}\) and \(\Mpro_{m}(d_1)\Mpro_{m}(d_2)=\Mpro_{m}(d_1+d_2)\).
The matrix that we obtain has the same form as the initial layer matrices, but with a total thicknes \(d_{tot}=d_1+d_2\).

Layers with different permittivity tensors can also be combined. At the boundary, interface matrices of the two layers combine as
\begin{equation}
\Mint_{mv}\Mint_{vn}
=
\Mint_{mn},\label{Eq:prodMint}
\end{equation}
to provide the usual interface matrix between \(m\) and \(n\).

The use of layer matrices with an interface to vacuum simplifies the analytical modelling, in particular because the vacuum is lossless and isotropic.

We have verified that layer matrices can be used quite efficiently to predict reflection and transmission at an interface, total internal reflection and Brewster's angle.

\subsection{Homogeneous multilayer materials }

\label{Sec:layerperlayer}

\begin{figure}
\includegraphics[width=\linewidth]{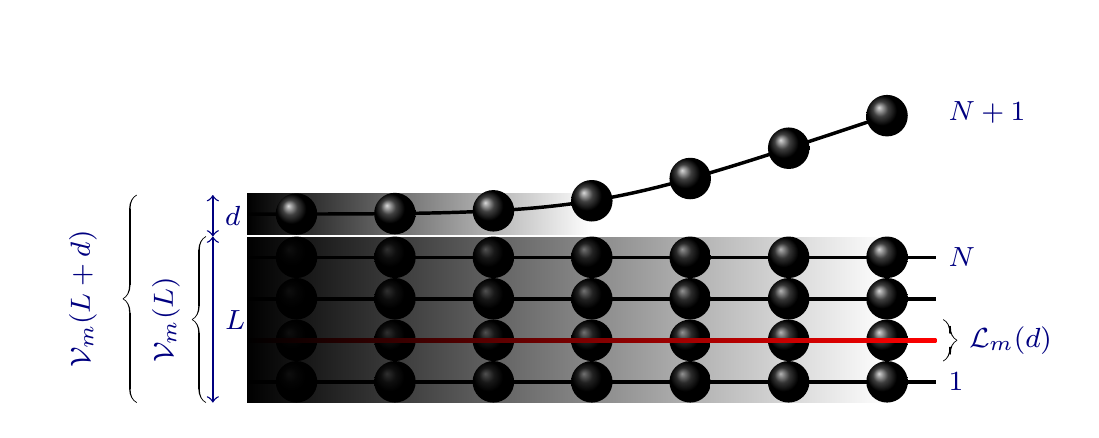}
\caption{Adding a 2D material on stacked layers. Illustration of volume properties of a continuous medium on the left, with the transfer matrix \(\MV_{m}\), and of a discrete medium on the right, with the single layer matrix \(\ML_{m}\). The red line indicates surface currents associated to one layer.}
\label{Fig:stackthelayers}
\end{figure}

To describe propagation in a bulk material, we could divide it in layers and stack them step by step as shown on Fig.~\ref{Fig:stackthelayers}.
Considering an arbitrary number \(N\) of layers in the volume, we can write the volume matrix \(\MV_m\) depending on the thickness \(L=Nd\), as
\begin{equation}
\MV_m(Nd) = \left[\ML_{m}(d)\right]^N
= \left[\Mint_{vm}
\Mpro_{m}(d)
\Mint_{mv}\right]^N
\stackrel{(\ref{Eq:prodI1I2})}{=}
\ML_m(Nd).
\end{equation}
The impact of layer \(N+1\) is therefore given by
\begin{equation}
 \frac{\delta\MV_{m}}{\delta{L}}
 = \frac{\MV_{m}(L+d)-\MV_{m}(L)}{d}
  = \frac{\ML_{m}(d)-I}{d}\,\MV_{m}(L).
 \label{Eq:derVm}
 \label{Eq:deltaVmdeltaL}
\end{equation}
If the medium is continuous, the thickness \(d\) of a layer is a continuously varying parameter. It can therefore be chosen arbitrarily small, which is equivalent to divide the continuous material in a very high number or layers. We can therefore consider the limit \(d\rightarrow0\), so that, using (\ref{Eq:LmDetaileProd}) and \(\Phi=k_z^m d\),  (\ref{Eq:deltaVmdeltaL}) becomes
\begin{eqnarray}
\frac{\drm\,\MV_{m}}{\drm{L}}
&=& \lim_{d\rightarrow0} \frac{\delta\MV_{m}}{\delta{L}} = K_{m} \MV_{m},\label{Eq:diffeqV}\\
K_{m}&=& \irm{}k_{z}^{m}\Mint_{vm}
    \matr{-1}{0}{0}{1}\Mint_{mv},
\label{Eq:Kmlayer}
\end{eqnarray}
as \(\lim_{d\rightarrow0} \left(\exp(\pm\irm{}k_{z}^{m}d)-1\right)/d =\pm\irm{}k_{z}^{m}\).
The matrix \(K_{m}\) contains all the information that is needed to build the volume layer. As it is obtained in the limit of small \(d\), it can be calculated from~(\ref{Eq:LayerMatrix}) or from its first order Taylor expansion in \(d\) or~\(\Phi\), showing that this first order expansion contains also all the physics of the system.
In this limit,
\begin{equation}
    \ML_{m}(d) = I + d
    K_m.\label{Eq:LinLayerMatrix}
\end{equation}
This last expression is interesting as it can be computed either from the continuous medium approach, or from the complex phase jump conditions at the interface. Indeed, it was shown in~\cite{majerus_electrodynamics_2018} that to the first order in \(d\) the thin film model and the interface model coincide and then
\begin{equation}
\ML_{m}=\Mint_{vm}\Mpro_{m}(d)\Mint_{mv}
= \Mpro_{v}(d/2) \mathcal{S}_{vv} \Mpro_{v}(d/2),
\label{Eq:LMcontvsdiscr}
\end{equation}
where the interface matrix with surface currents \(\mathcal{S}_{vv}\) is built from the reflection and transmission coefficients in (\ref{Eq:1overt}) and (\ref{Eq:rovert}) using the matrix expression in~(\ref{Eq:trmat}), in the case where \(a\) and~\(b\) correspond to vacuum. According to the {\ALM} in~\cite{majerus_electrodynamics_2018}, equation (\ref{Eq:LMcontvsdiscr}) holds when
\begin{eqnarray}
\frac{\chi_{x}^s}{d}&=&\frac{\varepsilon_x^m}{\varepsilon_0} - 1 ,\label{Eq:epschix}\\
\frac{\chi_{y}^s}{d}&=&\frac{\varepsilon_y^m}{\varepsilon_0} - 1,\label{Eq:epschiy}\\
\frac{\chi_{z}^s}{d}&=&\frac{\varepsilon_z^m-\varepsilon_0}{\varepsilon_z^m}
\stackrel{(\ref{Eq:xizs})}{=}\frac{\xi_{z}^s}{d}
.\label{Eq:epschiz}
\end{eqnarray}
We can therefore build \(K_m\) in two different ways, using the two forms of (\ref{Eq:LMcontvsdiscr}). This provides a direct analytical link between the anisotropic volume layer in [Fig.~\ref{Fig:sc-tf-lay}(III)] and the surface layer [Fig.~\ref{Fig:sc-tf-lay}(IV)]. An important consequence of our definition of a layer, as a dipolar distribution surrounded by vacuum is that the parameters appearing in \(K_m\) are independent from the effective surrounding media (\(a\) and  \(b\)) on Fig.~\ref{Fig:sc-tf-lay}. This was not the case in the {\AIM}.

 As a first step to our goal, which is to model heterogeneous stacked 2D materials, we will apply the approach of the previous section  in the approximation of the homogeneous continuous volume, where \(d\) can be chosen arbitrarily small. In this limit,
Eq.~(\ref{Eq:diffeqV}) is solved using matrix exponentials. As shown in \suprefsec{Sec:matrexp} we get
\begin{equation}
\MV_{m}(L)=\erm^{K_m L}\MV_{m}(0)
= \Mint_{vm}\matr{\erm^{-\irm{}k_z^mL}}
                 {0}{0}{\erm^{\irm{}k_z^mL}}
                 \Mint_{mv}.
\label{Eq:Vmint}
\end{equation}
Although this result is not surprising, as it corresponds to a layer matrix \(\ML_{m}(L)\), it is very instructive to demonstrate that \(K_m\) corresponding to a microscopic thickness \(d\rightarrow0\), is sufficient to describe an homogeneous material of macroscopic thickness \(L\). It allows to retrieve the propagation matrix and the interface matrices. This will be used in Sec.~\ref{Sec:hybridmat} to build the propagation matrix of an heterogeneous material and identify its effective permittivity tensor.

\subsection{Precision needed to distinguish the continuous and discrete approaches}
\label{Sec:discretemedium}

The equivalence between the continuous volume and the discrete layer approaches depicted respectively on pannels (III) and (IV) of Fig.~\ref{Fig:sc-tf-lay} is effective when \(d\rightarrow0\).

However, even if the thickness of 2D materials is generally negligible, we would like to calculate the precision that is needed on complex phase shift measurements to distinguish between the continuous and the discrete approaches when \(d=d_m\), the effective thickness of a 2D layer.

To this end, we will compare the matrix obtained in (\ref{Eq:Vmint}) to the one corresponding to stacked layers of thickness \(d_m\).
The number of layers to consider is given by \(L/d_m\), each layer being described by the matrix \(K_m\) using (\ref{Eq:LinLayerMatrix}) with \(d=d_m\).
The transfer matrix for the stacked layers is therefore given by
\begin{eqnarray}
\left(I+K_m d_m\right)^{L/d_m}.
\label{Eq:Vmprod}
\end{eqnarray}

From the analytical expressions of (\ref{Eq:Vmint}) and (\ref{Eq:Vmprod}), it is possible to calculate the difference on the complex phase accumulated over one layer (\(L=d_m\)), provided by the continuous and discrete approaches
\begin{equation}
\Delta\Phi(d_m) = {k_z^m d_m} - \ln\left(1+\irm{k_z^m d_m}\right)/\irm.
\end{equation}

By expanding \(\Delta\Phi\) in Taylor series, we demonstrate (see~\suprefsec{Sec:errordiscr}) that this error is bounded by
\begin{equation}
\left\vert\Delta\Phi\right\vert
< \frac{3}{2}\frac{\left\vert{k_z^m}\right\vert^2d_m^2}
       {\left(1-\mathrm{Im}[k_z^m] d_m\right)^3},
\end{equation}
in a lossy material (\(\mathrm{Im}[k_z^m]>0\)).

To get an estimate of this error bound, we consider a material with thickness 0.5\,nm, at a wavelength of 1000\,nm and a complex refractive index on the order of~\(1\), both for the real and imaginary parts. This leads to \(\vert{k_z^m}\vert{}d_m<\vert{n}\vert{}k_0d_m=2\cdot2\pi\cdot0.5/1000\approx6\cdot10^{-3}\), that is an absolute error
\(\left\vert\Delta\Phi\right\vert < 6\cdot10^{-5} \), and a relative error lower than \(0.01\).
For an effective refractive index of \(10\), this value would be reduced to \(0.1\).

At the moment, this error is small in view of the precision of the measurements performed on single-layer materials, but it could become important, especially at shorter wavelengths.

\section{Link between the existing models and our new anisotropic layer model}
\label{Sec:layertoBC}
\label{Sec:LayerfromBC}

\begin{figure}
\centering
\includegraphics[width=0.9\linewidth]{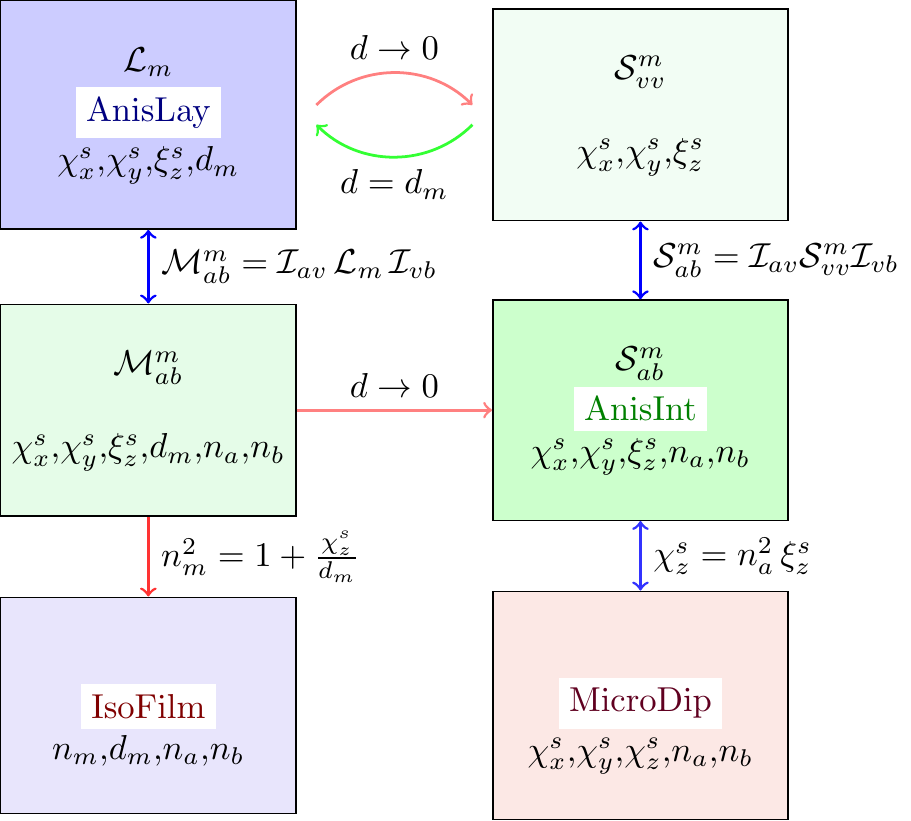}
\caption{Relations between the different models appearing in Tab.~\ref{Tab:sc-tf-lay}. Red arrows indicate loss of information. The green arrow requests additional information. Unnamed models are intermediate steps to link our new model to the existing ones.
Transfer matrices \(\Mint_{pq}\), \(\mathcal{S}_{vv}\) and \(\mathcal{L}_m\) are defined in the text. The parameters appearing at the bottom of each box denote the quantities used in the transfer matrices and/or the TE and TM reflection coefficients, and hence in the ellipsometric parameters.}
\label{Fig:diag_models}
\end{figure}

In Sec.~\ref{Sec:interfacematrix} we have presented the existing layer models that are described in terms of a complex phase shift induced at a boundary interface containing a bidimensional material. Then, in Sec.~\ref{Sec:layermodel} we have introduced our {\ALM} that takes into account the effective thickness of one layer of 2D material.

In order to compare these models, either from a theoretical or an experimental point of view, we should either calculate the phase shift induced at the interface from the layer matrix or calculate the layer matrix from the phase jump at the interface.

Our goal is to understand the relations between the different models in Tab.~\ref{Tab:sc-tf-lay}. The final result is shown on Fig.~\ref{Fig:diag_models}, where models on the left have a thickness \(d=d_m\) and those on the right are taken in the limit \(d\rightarrow0\). Red arrows between models correspond to loss of information, which is, for example, the case when going from an anisotropic model to an isotropic one, as in this case birefringence information is lost.

When presenting the {\AIM}, we have seen that the surface polarization terms are associated to the quantities \(\varphi_{ab}\) and \(\psi_{ab}\) in Eqs.~(\ref{Eq:BC:x}) and (\ref{Eq:BC:yz}). We will show here that these functions can be calculated from the {\ALM} (\ref{Eq:LayerMatrix}) with the energy conservative expansion of \(\exp(\irm\phi)\) described in appendix~(\ref{Eq:ConsApproxExp}) that is rewritten to separate the even and the odd parts in \(\Phi\)
\begin{eqnarray}
\cos(\Phi)+\irm\sin(\Phi)
&\approx&\frac{4-\Phi^2}{4+\Phi^2}
+\irm\frac{4\Phi}{4+\Phi^2}.
\end{eqnarray}
This provides (see~\suprefsec{Sec:TRratios}) the transmitance and reflectance:
\begin{eqnarray}
T_L
   &=&1+\frac{\irm\Phi_m/\alpha_{vm}}{2-\irm\Phi_m/\alpha_{vm}}
    +\frac{\irm\alpha_{vm}\Phi_m}{2-\irm\alpha_{vm}\Phi_m},\label{Eq:TLphialpha}\\
R_L
&=&\frac{\irm\alpha_{vm}\Phi_m}
         {2-\irm\alpha_{vm}\Phi_m}
         -
         \frac{\irm\Phi_m/\alpha_{vm}}
         {2-\irm\Phi_m/\alpha_{vm}}\label{Eq:RLphialpha}
         ,
\end{eqnarray}
from which one can calculate \begin{eqnarray}
T_{L}\left(\frac{1}{T_{L}}+\frac{R_{L}}{T_{L}}-1\right)
&=&-\frac{2\irm\Phi_m/\alpha_{vm}}
         {2-\irm\Phi_m/\alpha_{vm}}
         \label{Eq:1pRmTO2},\\
T_{L}\left(\frac{1}{T_{L}}-\frac{R_{L}}{T_{L}}-1\right)&=&-\frac{2\irm\alpha_{vm}\Phi_m}
         {2-\irm\alpha_{vm}\Phi_m}
         \label{Eq:1MDMTO2}.
\end{eqnarray}

Now comes the interesting physical interpretation. We expect that in the limit of infinitely thin layers, the right-hand side of these last equations will provide the non-continuity conditions of the electric field and displacement appearing in (\ref{Eq:BC:x}) and (\ref{Eq:BC:yz}).

We first consider that the permittivity and the surface susceptibility are related by (\ref{Eq:epschix})--(\ref{Eq:epschiz}) so that in the limit of very small \(d_m\),
\(
d_m\frac{\varepsilon_x^m-\varepsilon_0}{\varepsilon_0}\approx\chi_{x}^{s}\), and
\(d_m\frac{\varepsilon_x^m-\varepsilon_0}{\varepsilon_x^m}\approx\xi_{z}^{s}\).

Alternatively, we can compute the interface matrix \(\mathcal{S}_{vv}^m\) from \(\ML_{m}\) using back propagation
\begin{equation}
\mathcal{S}_{vv}^m
=\Mpro_{v}(-d_m/2)\ML_m(d_m)\Mpro_{v}(-d_m/2),
\label{Eq:Svv}
\end{equation}
and we calculate \(T_S\pm{R_s}-1\) from this matrix to identify \(\varphi_{ab}\) and \(\psi_{ab}\) from the phase jump at the 2D interface (\ref{Eq:BC:x})-(\ref{Eq:BC:yz}).

Both approaches provide the same results (See \suprefsec{Sec:interfacematrix}).
This is important as it explains why the layer \(\ML(d_m)\) can be presented either as a continuous layer as shown on Fig.~\ref{Fig:sc-tf-lay}(III), or as current sheet surrounded by vacuum, as depicted on Fig.~\ref{Fig:sc-tf-lay}(IV). This dual character of the {\ALM} allows to connect volume parameters such as permittivity, and propagation constant to surface parameters such as surface susceptibilities.

As \(\varphi_{ab}\) and \(\psi_{ab}\) have different expressions for TE and TM configurations, they should be calculated separately. This provides
\begin{eqnarray}
\phiTE[m]_{vv}&=&0,\label{Eq:phiTEvvm}\\
\phiTM[m]_{vv}&=&k_z^v\chi^s_{x},\label{Eq:phiTMvvm}\\
\psiTE[m]_{vv}&=&\frac{k_0^2}{k_z^v}\chi^s_{y},\label{Eq:psiTEvvm}\\
\psiTM[m]_{vv}&=&\frac{k_x^2}{k_z^v}\xi^s _{z}\label{Eq:psiTMvvm},
\end{eqnarray}
 with \(\xi_{z}^s\) defined in (\ref{Eq:xizs}).

Importantly, in (\ref{Eq:phiTEvvm})--(\ref{Eq:psiTMvvm}) there is no coefficient involving response functions of the input or output material (see top boxes in Fig.~\ref{Fig:diag_models}).

In a practical experiment, data should be interpreted by including for the input layer \(a\) (resp. the substrate \(b\)), interface matrices \(\Mint_{av}\) (resp. \(\Mint_{vb}\)).
Note that we are using \(k_x\) here, which is defined at the input as \(k_x=k_0\,n_a\sin\theta_a\). This can introduce an apparent dependency on the refractive index of the input medium \(n_a\) and on the incidence angle in this medium \(\theta_a\), though the layer parameters are intrinsic.

The inverse problem of inferring the effective parameters of the layer (propagation constant, permittivity) from the phase jump at the 2D interface is an important step to describe heterogeneous multilayer systems.

However, it is not straightforward to go from (\ref{Eq:phiTEvvm})--(\ref{Eq:psiTMvvm}) to a
factorized version of \(\ML_m\) in terms of \(\Phi_m\) and \(\alpha\).
 The main difficulty here is that the three matrices in (\ref{Eq:LmProd}) depend on the effective permittivity of the layer. We will show that it is nonetheless possible to define effective propagation parameters using the eigenvalues of the \(K_m\) matrix.

Starting from boundary conditions (\ref{Eq:BC:x}) and (\ref{Eq:BC:yz}), we build the interface matrix \(\mathcal{S}_{vv}^m\) of a 2D layer surrounded by vacuum using (\ref{Eq:trmat}) and the symmetry of the system leading to \(r=r^\prime, t=t^\prime\).
As detailed in~\suprefsec{Sec:intfromBC}, we get
\begin{equation}
 \mathcal{S}_{vv}=I+\matr{
 -\irm\frac{\varphi_{vv}}{2t}
 -\irm\frac{\psi_{vv}}{2t}}
 {\irm\frac{\varphi_{vv}}{2t}
 -\irm\frac{\psi_{vv}}{2t}}
 {-\irm\frac{\varphi_{vv}}{2t}
 +\irm\frac{\psi_{vv}}{2t}}
 {
  \irm\frac{\varphi_{vv}}{2t}
 +\irm\frac{\psi_{vv}}{2t}}.
 \label{Eq:SvvO1}
\end{equation}
The layer matrix \(\ML_m\) is then built using (\ref{Eq:LMcontvsdiscr})
with an effective thickness \(d_m\) that can be measured through electron microscopy~\cite{song_complex_2019}, and a first order expansion of
\begin{equation}
\mathcal{P}_v(d_m)=I+\irm{k_z^m}d_m\matr{-1}{0}{0}{1}.
\end{equation}
The first-order layer matrix is
\begin{eqnarray}
\ML_m(d_m)
&=& I + \irm\matr{-A}{B}{-B}{A}d_m,\label{Eq:LmABBA}\\
A&=&\frac{\varphi_{vv}^m}{2td_m}+\frac{\psi_{vv}^m}{2td_m}+k_z^v,\\
B&=&\frac{\varphi_{vv}^m}{2td_m}-\frac{\psi_{vv}^m}{2td_m}.
\end{eqnarray}
Eq.~(\ref{Eq:LmABBA}) has a convenient form to calculate \(K_m\) from (\ref{Eq:LinLayerMatrix}), and the volume matrix \(\MV_m(L)=\exp(K_mL)=\ML_m(L)\). This thick layer matrix has the form of (\ref{Eq:LmDetaileProd}). Parameters \(\alpha\) and \(\beta=\Phi/L\) are obtained by calculating the eigenvalues and eigenvectors of \(K_m\).
The eigenvalues provide the propagation constant \(\beta\) in the layer, while, through \(\alpha\), the eigenvectors provide the reflection and transmission coefficients between vacuum and the layer effective medium
\begin{eqnarray}
\beta^2&=&A^2-B^2
={\left(\frac{\varphi_{vv}}{td_m}+k_z^v\right)\left(\frac{\psi_{vv}}{td_m}+k_z^v\right)},
\label{Eq:betaAB}\\
\alpha&=&\frac{\beta}{\sqrt{A+B}}
=\frac{\beta}{\sqrt{\frac{\varphi_{vv}}{td_m}+k_z^v}}.
\label{Eq:alphaAB}
\end{eqnarray}
 By using the expressions (\ref{Eq:phiTEvvm})--(\ref{Eq:psiTMvvm}) for a TE and a TM input wave, we get as expected \(\alpha=\alpha_{vm}\) and \(\beta=k_z^m\).

This confirms the validity of our approach on homogeneous media.

We are therefore able to go from the right to the left side of the diagram in Fig.~\ref{Fig:diag_models}, which means that we can build a layer model from any bidimensional model providing at the 2D interface the phase jumps  (through \(\varphi_{vv}\) and \(\psi_{vv}\)) and the effective thickness of the layer \(d_m\).

\section{Heterostructures of 2D materials and intrinsic surface quantities}
\label{Sec:hybridmat}

\begin{figure}
\centering\includegraphics[width=0.8\linewidth]{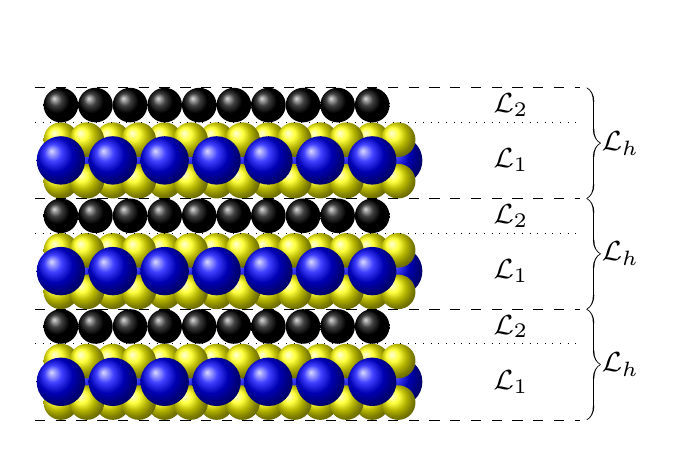}
\caption{Hybrid structure made of alternated layers of type \(1\) and~\(2\). The structure is modeled as a set of hybrid layers (\(h\)).}
\label{Fig:hybrid}
\end{figure}
It is now time to show the importance of the layer approach by providing effective parameters for an heterostructure made of stacked alternating layers of 2D materials.

In our {\ALM}, we consider two kinds of layers characterized by two different susceptibility tensors and thicknesses (Fig.~\ref{Fig:hybrid}).
The elementary layer of this hybrid material corresponds to the product of two single layers
\begin{eqnarray}
    \lefteqn{\ML_{h} = \ML_{1} \ML_{2}}\\
    &=& \left[I + \irm\matr{-A_1}{B_1}{-B_1}{A_1}d_1\right]\times\nonumber\\
    &&   \left[I + \irm\matr{-A_2}{B_2}{-B_2}{A_2}d_2\right],
\end{eqnarray}
that provides, at first order, a \(K_h\) matrix similar to (\ref{Eq:LmABBA}), with \(A=A_1d_1+A_2d_2\), \(B=B_1d_1+B_2d_2\) and \(d_h=d_1+d_2\). Using (\ref{Eq:betaAB})-(\ref{Eq:alphaAB}) this leads to
\begin{eqnarray}
\beta_{\TE}^2&=&\frac{d_1}{d_h}\beta_1^2+\frac{d_2}{d_h}\beta_2^2,\\
\beta_{\TM}^2&=&\sum_{i=1}^2\frac{d_i}{d_h}\beta_i^2\left(\frac{d_i}{d_h}+\frac{d_{3-i}}{d_h}\frac{\varepsilon_{x,{3-i}}}{\varepsilon_{x,i}}\right).
\end{eqnarray}
After expanding \(k_z^{x},\;x=v,1,2,h\) as in (\ref{Eq:kzTE}) and (\ref{Eq:kzTM}) we get the effective permitivities of the hybrid medium
\begin{eqnarray}
\varepsilon_x^h&=&\frac{d_1}{d_h}\varepsilon_{x,1}
                +\frac{d_2}{d_h}\varepsilon_{x,2},\label{Eq:epsxhybrid}\\
\varepsilon_y^h&=&\frac{d_1}{d_h}\varepsilon_{y,1}
                +\frac{d_2}{d_h}\varepsilon_{y,2},
                \label{Eq:epsyhybrid}\\
\frac{1}{\varepsilon_z^h}&=&
             \frac{d_1}{d_h}\frac{1}{\varepsilon_{z,1}}
                +\frac{d_2}{d_h}\frac{1}{\varepsilon_{z,2}}
                \label{Eq:epszhybrid}.
\end{eqnarray}
\subsection{Discussion}
\label{Sec:HybridDiscussion}

If we expand the in-plane permittivity in terms of in-plane susceptibility, (\ref{Eq:epsxhybrid}) and (\ref{Eq:epsyhybrid}) take the form
\begin{eqnarray}
1+\chi_h&=&\frac{d_1}{d_h}\left(1+\chi_1\right)
        +\frac{d_2}{d_h}\left(1+\chi_2\right),\\
d_h\chi_h&=&d_1\chi_1+d_2\chi_2,
\end{eqnarray}
showing that if the surface susceptibility is \(\chi^s=d\chi\), the total surface susceptibility is the sum of the individual surface susceptibilities, in which the thicknesses of the layers do not appear : \(\chi_h^s=\chi_1^s+\chi_2^s\).

We see that the order of the layers does not enter into account. This conclusion is valid within the error bound given in Sec.~\ref{Sec:discretemedium}.
It corresponds to the classical vision that on a subwavelength scale, dipole contributions can be summed to model the material response.

We see here that the total dipole contribution is handled separately from the total thickness. Though we can consider pure surface contributions from the dipole response of the 2D materials, we cannot neglect the thickness of the individual layers.

However, for the out-of-plane component, it is convenient to use the displacement susceptibility \(\xi\) defined in (\ref{Eq:xizs}) so that (\ref{Eq:epszhybrid}) becomes
\begin{eqnarray}
\frac{1}{\varepsilon_0}-\frac{1}{\varepsilon_{z,h}}
&=&\frac{1}{\varepsilon_0}-\frac{d_1}{d_h}\frac{1}{\varepsilon_{z,1}}
    -\frac{d_2}{d_h}\frac{1}{\varepsilon_{z,2}}\label{Eq:chixyhybrid},\\
d_h\xi_h&=&d_1\xi_1+d_2\xi_2\label{Eq:xizhybrid}
.
\end{eqnarray}
Therefore, as we have defined in (\ref{Eq:xizs}) that \(\xi^s=d\,\xi\),
\begin{equation}
\xi_h^s=\xi_1^s+\xi_2^s
.
\end{equation}

An important conclusion from (\ref{Eq:epsxhybrid}) is that out-of-plane surface susceptibilities \(\chi_z^s\) do not sum in hybrid materials while the surface displacement susceptibilities \(\xi_z^s\) do.

\label{Sec:IntrinsicSurfaceQuantities}

Equations (\ref{Eq:chixyhybrid}) and (\ref{Eq:xizhybrid}) show that the displacement susceptibility \(\xi_z^s\) defined in (\ref{Eq:xizs}) is the surface quantity that should be used to describe the out-of-plane response of a surface material, while \(\chi_{x}^s\) and \(\chi_{y}^s\) are used for the in-plane response.

To summarize previous results, the intrinsic surface quantities are defined through
\begin{eqnarray}
\chi_{x}^{m,s}&=&\frac{\mathcal{P}_x}{\varepsilon_0E_x^m}
                =\frac{\mathcal{P}_x}{\varepsilon_0E_x^t}
                =d_m\chi_{x}^{m} = d_m\frac{\varepsilon_{x}^m-\varepsilon_0}{\varepsilon_0},\label{Eq:chixxsummary}\\
\xi_{z}^{m,s}&=&\frac{\mathcal{P}_z}{D_z^m}
              = \frac{\mathcal{P}_z}{D_z^t}
              = d_m\xi_{z}^{m} = d_m\frac{\varepsilon_{z}^{m}-\varepsilon_0}{\varepsilon_{z}^{m}},\label{Eq:xizzsummary}
\end{eqnarray}
with \(t\) denoting the transmitted fields and where the \(y\) component is obtained by the substitution \(x\rightarrow{y}\) in~(\ref{Eq:chixxsummary}).

These quantities are intrinsic in that they do not depend on the parameters of the surrounding layers. To clearly understand the meaning of this, remember that we have defined a layer as being surrounded by vacuum.
In practice, when we  evaluate the surface quantities of a layer \(m\) on a substrate \(b\), we could do so based on the field in medium \(m\) [seen as a continuous medium, Fig.~\ref{Fig:sc-tf-lay}(III)]; in medium \(b\); or in the vacuum when the layer is seen as a current sheet surrounded by vacuum [see Fig.~\ref{Fig:sc-tf-lay}(IV)].

The field and displacement components that appear in (\ref{Eq:chixxsummary}) and (\ref{Eq:xizzsummary}) are continuous across a surface without surface currents.
Indeed, although we consider an interface with surface currents, we expect the phase shift induced by the currents to be small, as discussed in~\cite{Li_Heinz_18} and it can be neglected to calculate the field in the layer. Under this approximation, the tangential part of the electric field \(\vec{E}_{tg}\) (two components), and the normal component of the displacement field \(D_z\) (one component), are continuous across the interface.
This means that we can decompose the interface matrix between medium \(m\) and substrate according to \(\Mint_{mb} = \Mint_{mv}\Mint_{vb}\), and assume that the three components mentioned above are the same in \(m\),  \(v\) and \(b\), confirming that the surface quantities \(\chi_{x}^{m,s}\), \(\chi_{y}^{m,s}\) and \(\xi_{z}^{m,s}\) are intrinsic.

These three quantities can be summed directly when stacking layers.
This is especially useful to build an heterogeneous multilayer material, as the number of layers is limited. We remind here that calculations in Sec.~\ref{Sec:hybridmat} were performed in the first order limit and their precision is limited as estimated in Sec.~\ref{Sec:discretemedium}.

\section{Ellipsometry for single- and multi-layer systems}
\label{Sec:fitting}

Experimental data about optical properties of 2D materials at different angles are very limited in the literature. This is probably due to the difficulty to record them with classical setups, as described in~\cite{majerus_electrodynamics_2018,adamson_non-standard_2019}. To circumvent this difficulty, Xu~\cite{xu_complex_2015} proposed an experimental configuration in which a 2D material is characterized in two steps: (i) on a polymer substrate; (ii) with the same polymer added on top, so that the 2D material is immersed in the polymer.
As a test of our new {\ALM}, we will investigate if it is compatible with these experimental data.

The data in~\cite{xu_optical_2021} are taken at a wavelength of 633\,nm on a graphene and an \chem{MoS_2} sample immersed in a polymer with refractive index \(n_p=1.4233\) at this wavelength. Four quantities are reported: the TE and TM reflectance, and the \(\Psi\) and \(\Delta\) ellipsometric parameters.
The {\MDM} is fitted to these data with a good agreement. Analytical curves for this model are reproduced for graphene on Figs.~\ref{Fig:fitgraphene}.

As shown on Fig.~\ref{Fig:diag_models} the different models do not use the same parameter for the out-of-plane polarization component. As the {\AIM} and the {\MDM} are both interface models, we will first use them to link \(\chi_{z}^s\) and \(\xi_z^s\).

In \suprefsec{Sec:microscopic_vs_lay}, we compare the analytical expressions of the two models, and show that they are equivalent if, in matrix \(\mathcal{S}_{ab}^m = \Mint_{av}\mathcal{S}_{vv}^m\Mint_{vb}\), we set
\begin{equation}
\chi_z^s=n_a^2\,\xi_z^s
.\label{Eq:xichiMDM}
\end{equation}

Starting from the surface susceptibilities published in~\cite{xu_optical_2021}, and reproduced in Tab.~\ref{Tab:fittingparams}, we can therefore calculate the line for the {\AIM} in this table.
As both models are described by the same analytical expression, they are represented together in Fig.~\ref{Fig:fitgraphene}, where the TE and TM reflectance curves with respect to the input angle appear on the left, and the ellipsometric data appear on the right.

If the distinction that we made between the layer and the interface model is significant, we expect to find a difference in the reflectance and ellipsometric curves of the {\ALM} plotted with the parameters of the {\AIM} (line 2 in Tab.~\ref{Tab:fittingparams}).
Fig.~\ref{Fig:fitgraphene} shows that this is indeed the case for the TM reflectance curve past the pseudo-Brewster angle. However, the discrepancy for the TE mode is quite limited.
Note that for the purpose of plotting data, the {\(\mathcal{L}_m\) matrix cannot be used directly, as it does not take into account the polymer surrounding the bidimensional layer. To this end, we build
\begin{equation}
\mathcal{M}_{aa}^m=\Mint_{av} \ML_{m}(d_m) \Mint_{va}
\label{Eq:MabLm}.
\end{equation}
Then, from the general shape of a transfer matrix \(\mathcal{M}\) [Eq.~(\ref{Eq:trmat})], we extract the reflection coefficient as a ratio of matrix components \(r=\mathcal{M}_{21}/\mathcal{M}_{11}\). Doing so for TE and TM modes allows to retrieve the reflectance curves \(R(\theta)=\vert{r(\theta)}\vert^2\) and the ellipsometric ratio \(\rho={r}^{\TM}/{r}^{\TE}=\tan\Psi\erm^{\irm\Delta}\), from which we get the ellipsometric parameters
\(\Psi\) and \(\Delta\).

Fig.~\ref{Fig:fitgraphene} confirms our expectation that different models will provide different retrieved parameters for the same set of data, raising the question of which are the intrinsic parameters. When comparing the {\AIM} and the {\MDM}, we have found equation (\ref{Eq:xichiMDM}) that relates both. It would be nice to find a similar link between the {\ALM} and the {\AIM}. We could not do so analytically, and resorted to numerical optimisation to match the curves of the {\AIM} with those of the {\ALM} by properly choosing the parameters.
With the values reported on line ``AnisLay'' in Tab.~\ref{Tab:fittingparams}, we can match the two models almost perfectly, as reported on Fig.~\ref{Fig:fitgraphene}, where the optimally matching curve is identified by diamonds.

\begin{figure}
\setlength{\tabcolsep}{0pt}
\renewcommand{\arraystretch}{0.5}
\begin{tabular}{cc}
 \includegraphics[width=0.48\linewidth]{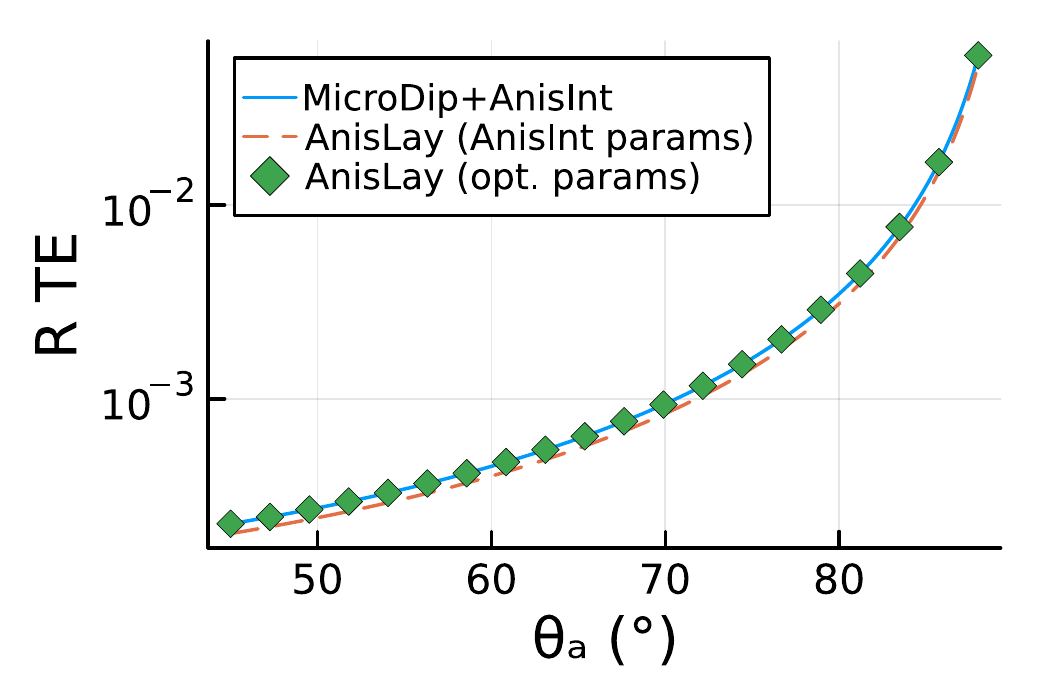}&
 \includegraphics[width=0.48\linewidth]{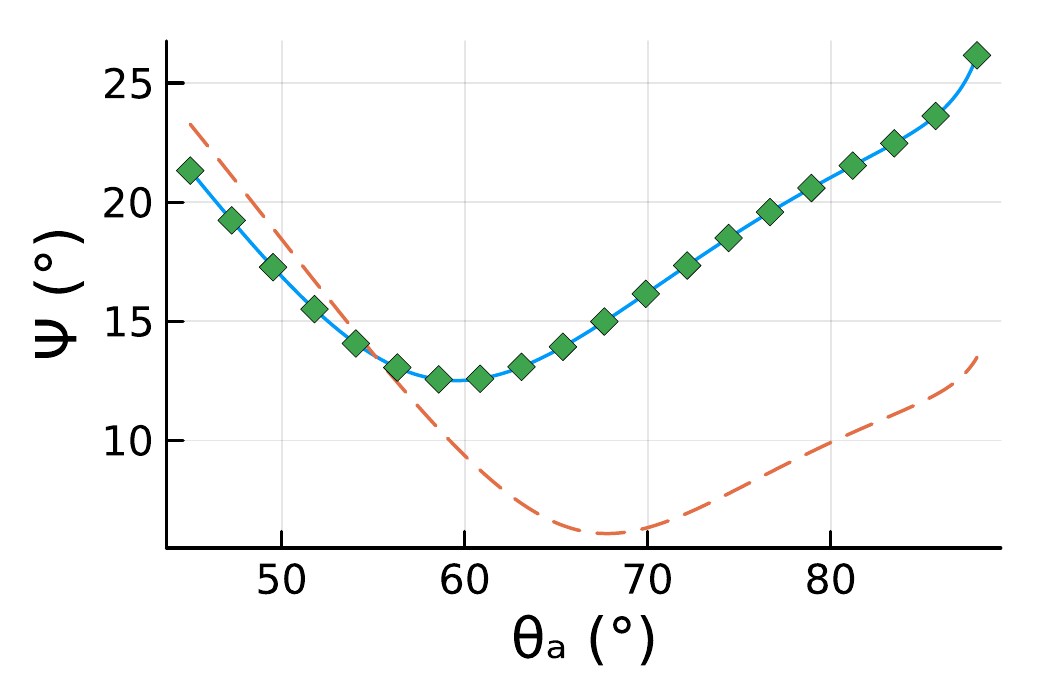}\\
 \includegraphics[width=0.48\linewidth]{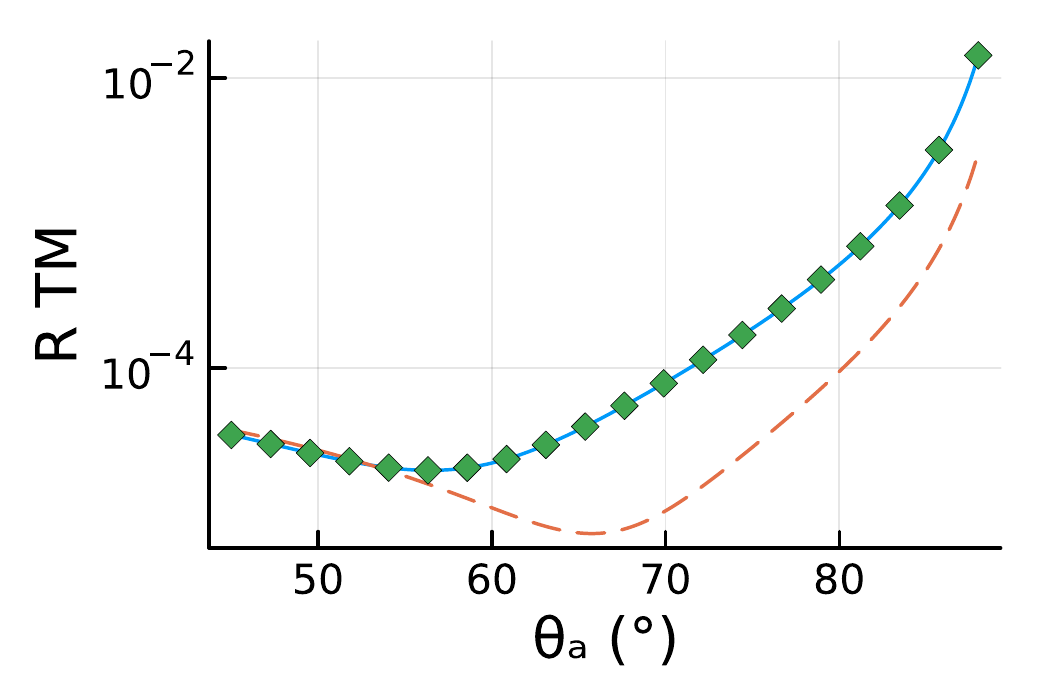}&
 \includegraphics[width=0.48\linewidth]{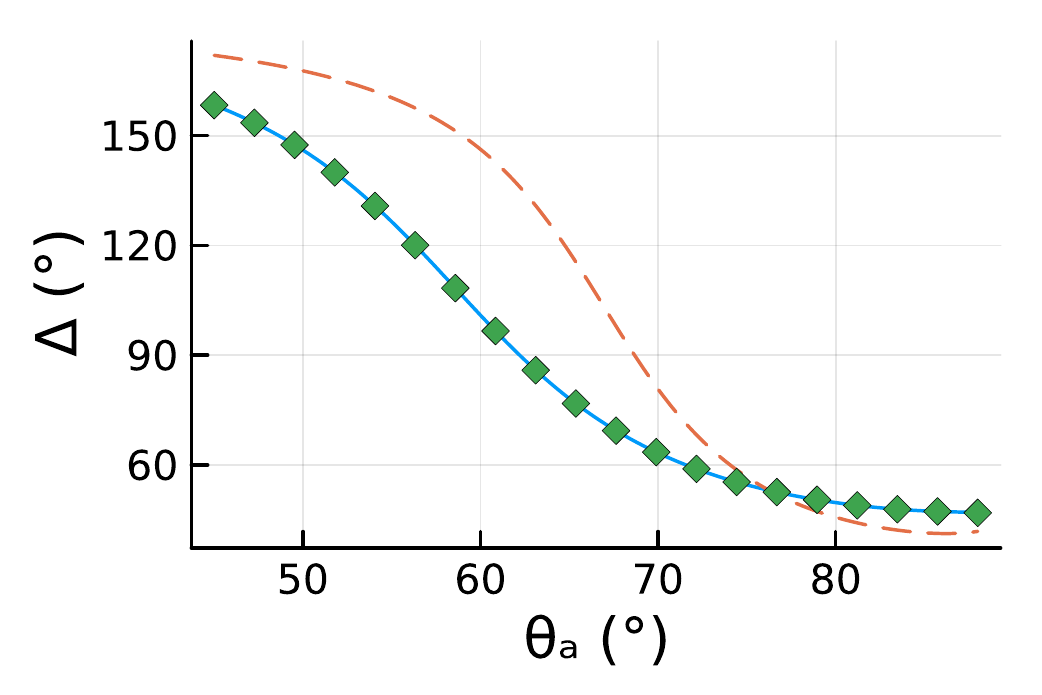}\\
\end{tabular}
\caption{Graphene reflectance and ellipsometric curves at a wavelength of 633\,nm. Our new model is compared to the interface response of the existing models. No experimental data are plotted here. Acronyms corresponds to those of Tab.~\ref{Tab:sc-tf-lay}. {\MDM} and {\AIM} (blue line) describe the interface and provide the same analytical expressions if (\ref{Eq:xichiMDM}) is applied. The {\ALM} plotted with the parameters of the {\AIM} differs strongly from it (red dashed). An optimal tuning of the {\ALM} parameters provides almost perfectly matching curves (green diamonds).}
\label{Fig:fitgraphene}
\end{figure}

The difference between the {\ALM} and the {\AIM} comes from the additional propagation in vacuum in {\ALM}, and mainly affects the real part of the susceptibilities, and of the in-plane permittivity. We should note that this additional thickness has an impact on both the real and imaginary parts of the out-of-plane permittivity.

\begin{table}
\caption{Comparison of the parameters used in the different models to reproduce the reflectance and the ellipsometric curves used in~\cite{xu_optical_2021} for fitting the experimental data for graphene at 633\,nm. For the {\AIM} \&\ {\ALM}, \(\chi_{z}^s=n_a^2\xi_z\). For all models, \(\varepsilon_i=\varepsilon_0(1+\chi_i^s/d_m)\), \((i=x,y,z)\). Exponents \(r\) and \(i\) denote real and imaginary parts. For graphene \(d_m=0.334\)\,nm.}
\label{Tab:fittingparams}
\footnotesize
\begin{tabular}{l|rr|rr|rr}
Model&\(\chi_{x}^{s,r}\)&\(\chi_{x}^{s,i}\)
     &\(\chi_{z}^{s,r}\)&\(\chi_{z}^{s,i}\)
     &\(\xi_{z}^{s,r}\)&\(\xi_{z}^{s,i}\)\\
&(nm)&(nm)&(nm)&(nm)&(nm)&(nm)\\
\hline
MicroDip&1.7&2.58&0.60&0.11&\(\cdot\)&\(\cdot\)\\
AnisInt&1.7&2.58&0.60&0.11&0.30&0.056\\
AnisLay&2.04&2.58&0.94&0.11&0.47&0.056
\end{tabular}

\strut

\begin{tabular}{l|rr|rr}
Model&\(\varepsilon_{x}^r\)&\(\varepsilon_{x}^i\)
     &\(\varepsilon_{z}^r\)&\(\varepsilon_{z}^i\)\\
\hline
AnisInt&6.09&7.73&2.63&4.35\\
AnisLay&7.12&7.73&-2.10&0.86
\end{tabular}
\end{table}

The same analysis on \chem{MoS_2} leads to the curves in Fig.~\ref{Fig:fitMoS2} and Tab.~\ref{Tab:fittingparamsMoS2}, with similar conclusions as for graphene.

\begin{figure}[ht!]
\setlength{\tabcolsep}{0pt}
\renewcommand{\arraystretch}{0.5}
\begin{tabular}{cc}
 \includegraphics[width=0.48\linewidth]{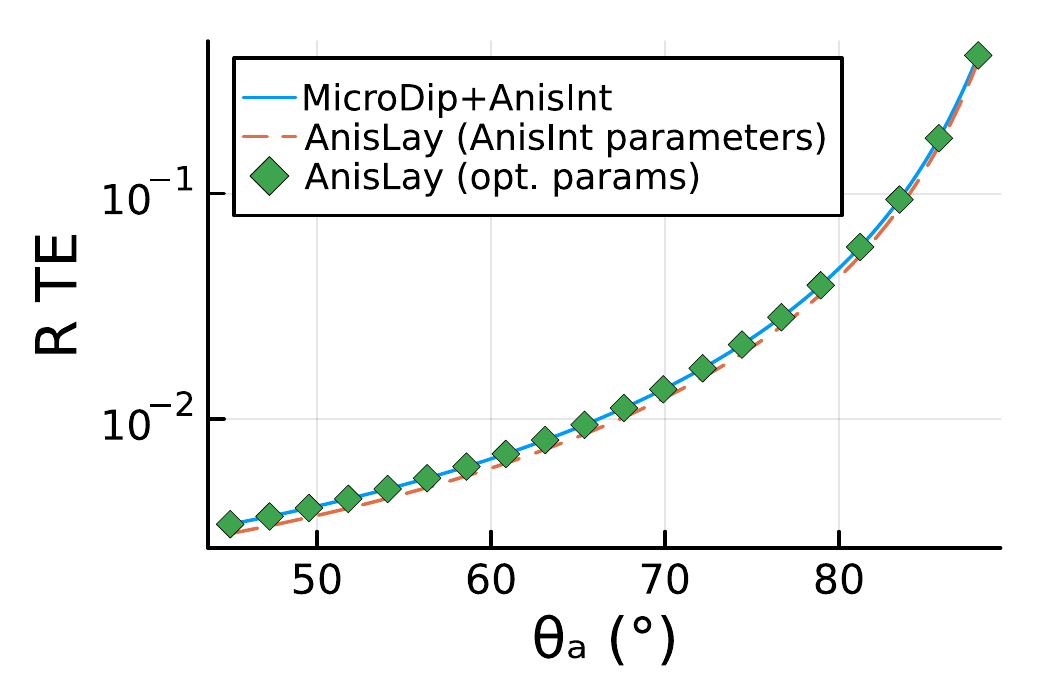}&
 \includegraphics[width=0.48\linewidth]{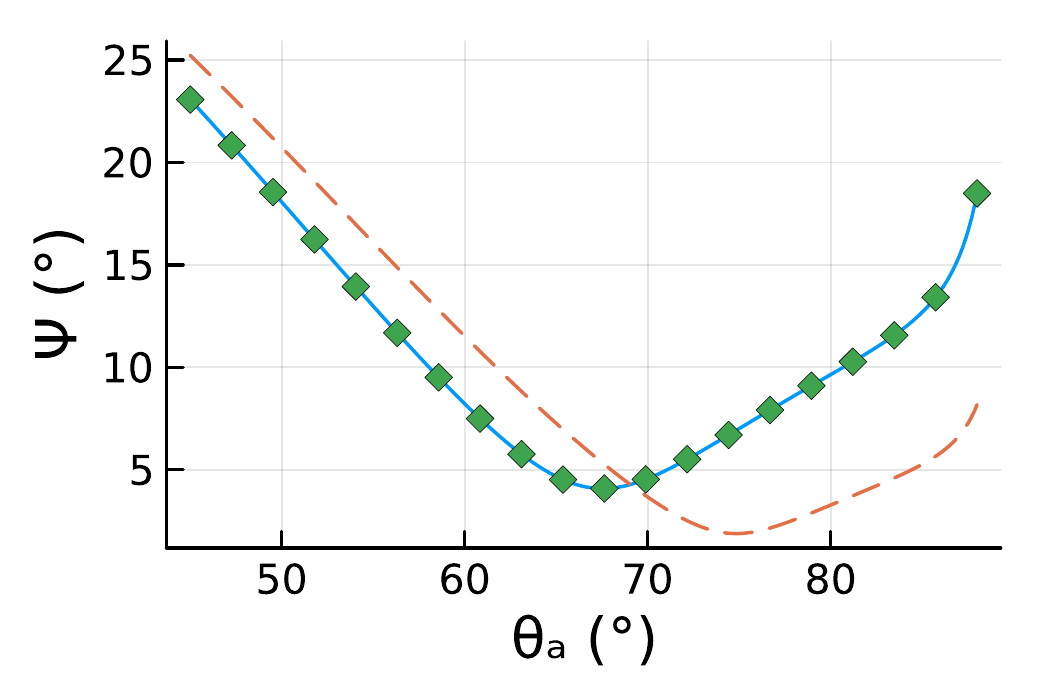}\\
 \includegraphics[width=0.48\linewidth]{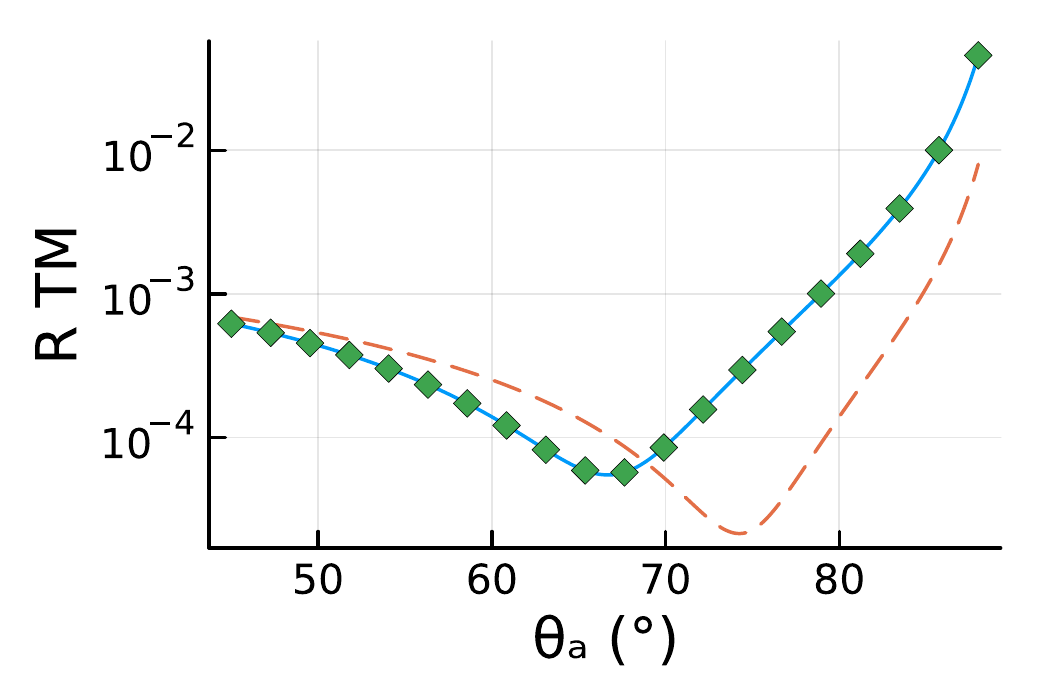}&
 \includegraphics[width=0.48\linewidth]{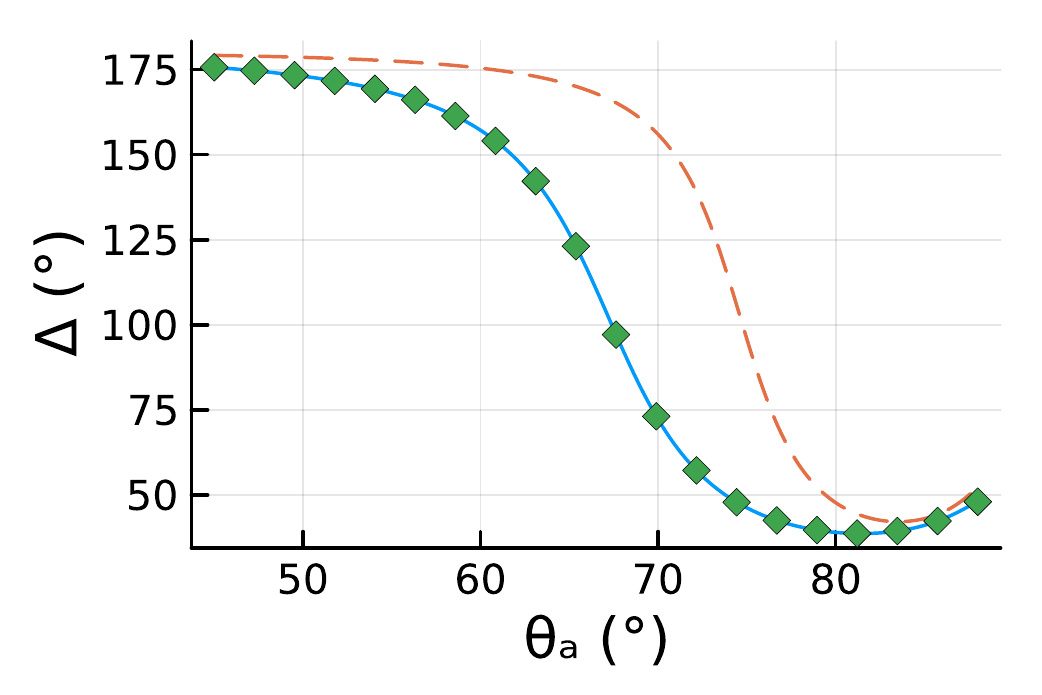}\\
\end{tabular}
\caption{\chem{MoS_2} reflectance and ellipsometric curves at a wavelength of 633\,nm. Our new model is compared to the interface response of the existing model. No experimental data are used here. Acronyms correspond to those of Tab.~\ref{Tab:sc-tf-lay}. {\MDM} and {\AIM} (blue line) describe the interface and provide the same analytical expressions if (\ref{Eq:xichiMDM}) is applied. The {\ALM} plotted with the parameters of the {\AIM} differs strongly from it (red dashed). An optimal tuning of the {\ALM} parameters provides almost perfectly matching curves (green diamonds).}
\label{Fig:fitMoS2}
\end{figure}

\begin{table}
\caption{Comparison of the parameters used in the different models to reproduce the reflectance and the ellipsometric curves used in~\cite{xu_optical_2021} for fitting the experimental data for \chem{MoS_2} at 633\,nm. Notations and parameters are the same as in Tab.~\ref{Tab:fittingparams}. For \chem{MoS_2}, \(d_m=0.631\)\,nm.}
\label{Tab:fittingparamsMoS2}
\footnotesize
\begin{tabular}{l|rr|rr|rr}
Model&\(\chi_{x}^{s,r}\)&\(\chi_{x}^{s,i}\)
     &\(\chi_{z}^{s,r}\)&\(\chi_{z}^{s,i}\)
     &\(\xi_{z}^{s,r}\)&\(\xi_{z}^{s,i}\)\\
&(nm)&(nm)&(nm)&(nm)&(nm)&(nm)\\
\hline
MicroDip&10.8&5.69&1.1&0.038&\(\cdot\)&\(\cdot\)\\
AnisInt&10.8&5.69&1.1&0.038&0.54&0.019\\
AnisLay&11.45&5.69&0.86&0.018&0.43&0.009\\
\end{tabular}

\strut

\begin{tabular}{l|rr|rr}
Model&\(\varepsilon_{x}^r\)&\(\varepsilon_{x}^i\)
     &\(\varepsilon_{z}^r\)&\(\varepsilon_{z}^i\)\\
\hline
AnisInt&18.12&9.02&6.64&1.39\\
AnisLay&19.15&9.02&3.13&0.14
\end{tabular}
\end{table}

Summarizing the previous results, the three models in Tab.~\ref{Tab:fittingparams} provide the same reflectance and ellipsometric parameters for different electromagnetic parameters. These models can therefore reproduce experimental results of ellipsometry with the same accuracy but with different susceptibilities and permittivities. However, as illustrated on Fig.~\ref{Fig:diag_models}, the newly proposed {\ALM} involves only intrinsic parameters in the description of the layer. Previous models provide therefore non-intrinsic parameters in Tab.~\ref{Tab:fittingparams}, which means that using the technique of~\cite{xu_optical_2021}, the retrieved susceptibilities of a given 2D material will vary if a different surrounding polymer is used for the measurement.

Moreover, as these models ignore the thickness of the 2D layer, they cannot be used directly to model multilayer systems.

The transfer matrix formalism accounts for multiple reflexions in stratified media and is therefore suited to the study of multilayer systems.

However, to build the matrix of the system, we must consider a model and a geometry compatible with the model.

Different possibilities are presented in Fig.~\ref{Fig:multilayer}.
Interface models consider a thickness \(d=0\) with a dipolar response concentrated at the interface. A N-layer interface model corresponds to the transfer matrix
\begin{equation}
\mathcal{M}_\mathrm{int}= \left(\mathcal{S}_{ab}^m\right)^N
\label{Eq:MintN}.
\end{equation}
The layer model, on the other hand provides
\begin{equation}
\mathcal{M}_\mathrm{lay}=\Mint_{av}\left(\mathcal{L}_{m}\right)^N\Mint_{vb}.
\label{Eq:MlayN}
\end{equation}
Interestingly, for the immersed configuration where \(I_{av}=I_{vb}^{-1}\), we have that
\begin{equation}
\left(\mathcal{S}_{ab}^m\right)^N
=
\left(\Mint_{av}\mathcal{S}_{vv}^m\Mint_{av}^{-1}\right)^N
=
\Mint_{av}\left(\mathcal{S}_{vv}^m\right)^N\Mint_{vb},
\label{Eq:MintNimmersed}
\end{equation}
which is similar to (\ref{Eq:MlayN}), with \(\mathcal{S}_{vv}\) taking the place of \(\mathcal{L}_m\). In this case, we also have that
\begin{equation}
\left(\mathcal{M}_{ab}^m\right)^N
=
\Mint_{av}\left(\mathcal{L}_{m}\right)^N\Mint_{vb}.
\end{equation}

Another common model used in the litterature considers that the dipolar response is concentrated in a current sheet interface, with vacuum added (usually on top) to take into account the real thickness (\(Nd_m\)). Here, we consider three configurations depicted on Fig.~\ref{Fig:multilayer}, with the current sheet interface at the bottom, in the middle or at the top of the vaccuum layer. These correspond to
\begin{equation}
\mathcal{M}_\mathrm{intvac,\eta}=
\Mint_{av}
\Mpro_{v}\left(\eta{}d_m\right)\left(\mathcal{S}_{vv}^m\right)^N
\Mpro_{v}\left((1-\eta)d_m\right)
\Mint_{vb},
\label{MintvacN}
\end{equation}
with \(\eta=0,1/2,1\) respectively.
From these matrices, we can again extract the reflection coefficient \(r=\mathcal{M}_{21}/\mathcal{M}_{11}\).

\label{Sec:MultilayerSyst}

\begin{figure}
\includegraphics[width=\linewidth]{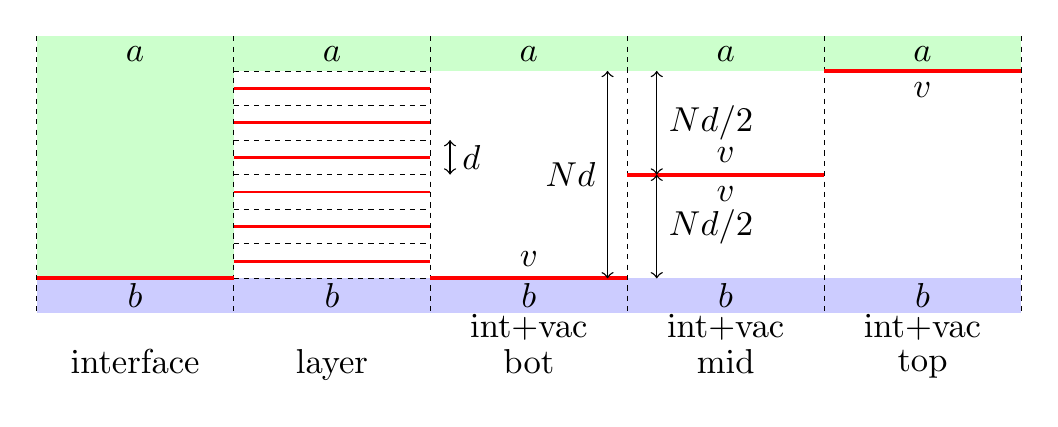}
\caption{Multilayer system. Schematic representation of the different models. \(a\) (green) and \(b\) (blue) denote surrouding media, \(v\) (white) represents vacuum. Red lines correspond to surface currents. The interface configurations ({\MDM} and {\AIM}) do not contain vacuum between the surface current and the surrounding material. In the layer model ({\ALM}) surface currents are separated by vacuum. In other representations, a single current sheet is considered, with vacuum on one or two sides to compensate for the physical thickness.
}
\label{Fig:multilayer}
\end{figure}

We evaluate the reflectance and ellipsometric curves predicted for the five geometries of Fig.~\ref{Fig:multilayer}, with 10 and 100 layers. For the interface model, we consider the parameters of the {\AIM} in Tab.~\ref{Tab:fittingparams} and Tab.~\ref{Tab:fittingparamsMoS2}, while for the other models we consider the intrinsic parameter reported in the last line of each table.

\begin{figure}
\setlength{\tabcolsep}{0pt}
\renewcommand{\arraystretch}{0.5}
\begin{tabular}{cc}
 \includegraphics[width=0.48\linewidth]{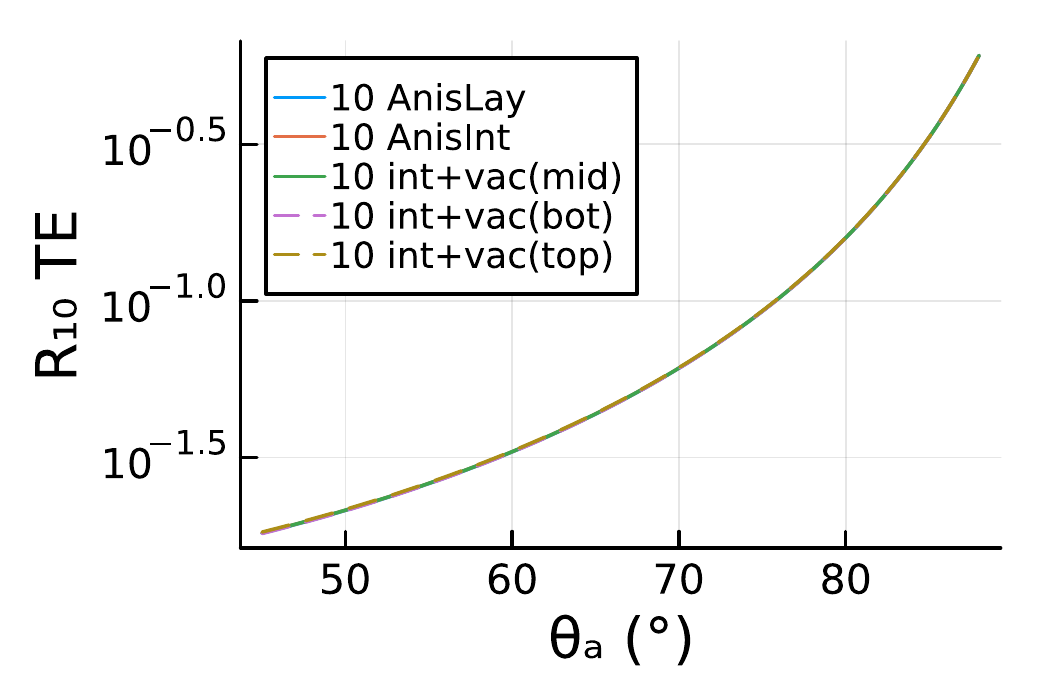}&
 \includegraphics[width=0.48\linewidth]{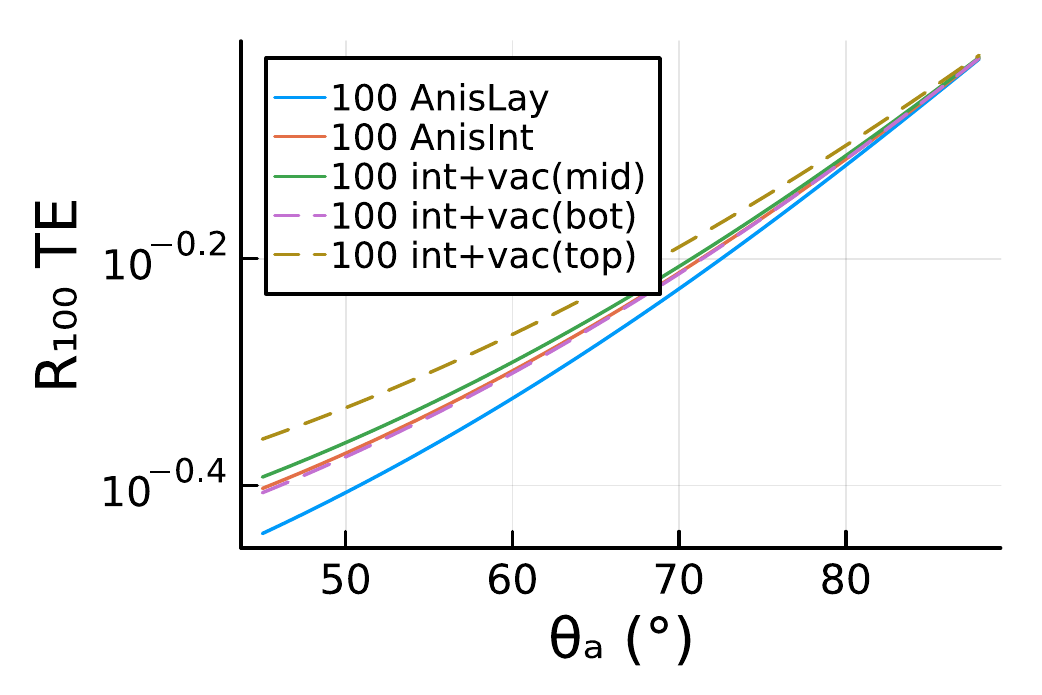}\\
 \includegraphics[width=0.48\linewidth]{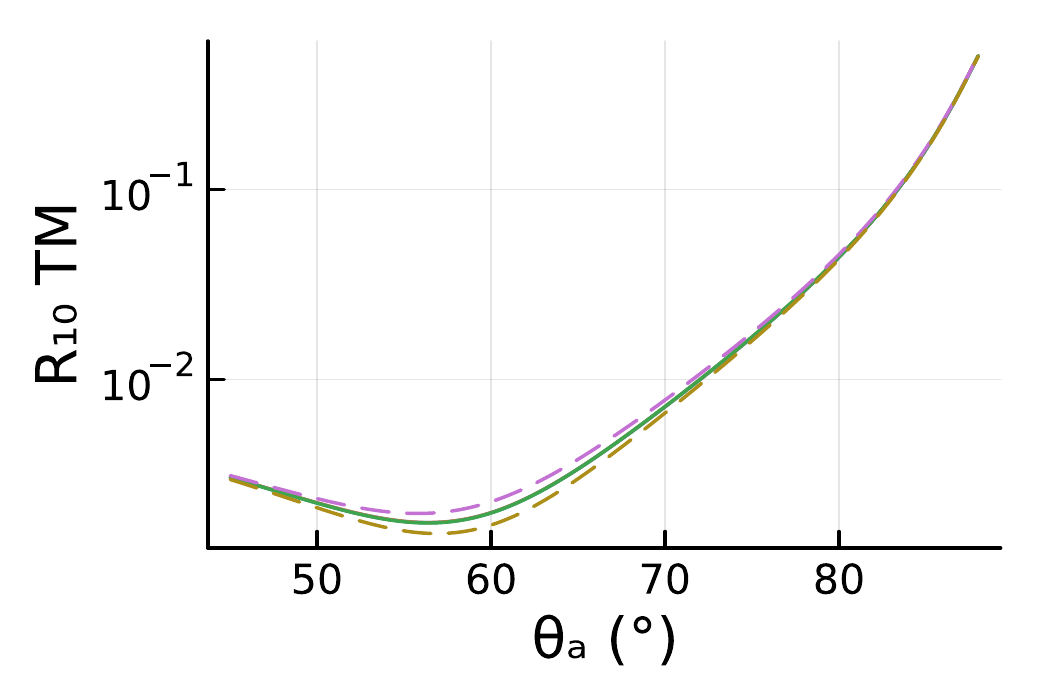}&
 \includegraphics[width=0.48\linewidth]{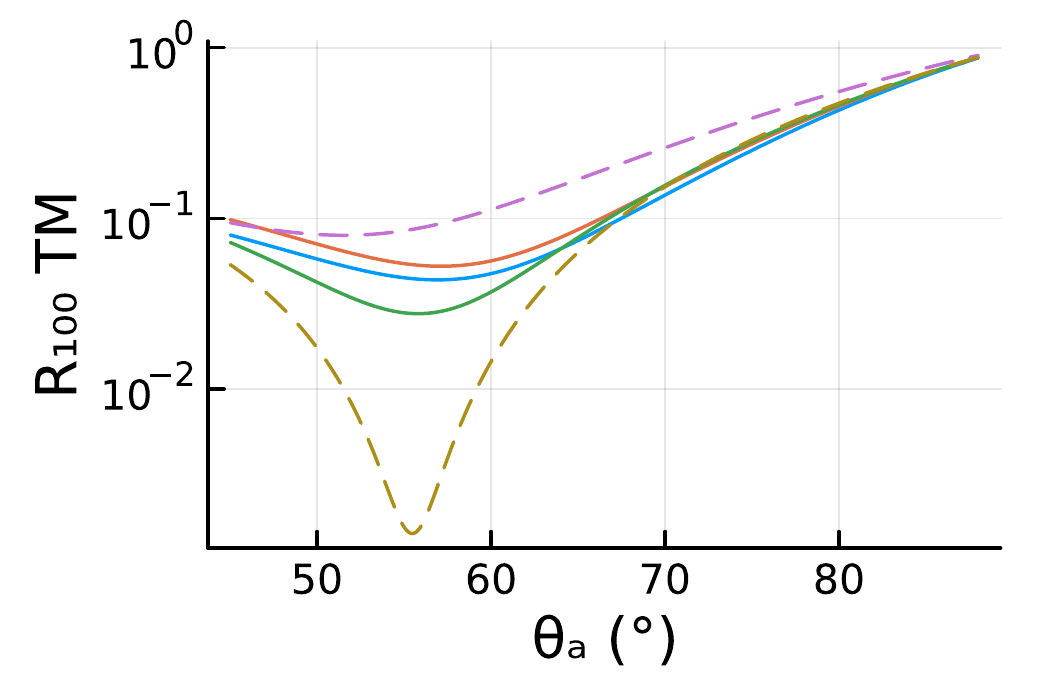}\\
\end{tabular}
\caption{Reflectance of multilayer systems of graphene sheets immersed in a polymer with refractive index 1.4233. The five geometries of Fig.~\ref{Fig:multilayer} are considered. Left: 10 layers. Right: 100 layers. Top, TE reflectance. Bottom, TM reflectance. }
\label{Fig:10L100L}
\end{figure}

\begin{figure}[ht!]
\setlength{\tabcolsep}{0pt}
\renewcommand{\arraystretch}{0.5}
\begin{tabular}{cc}
 \includegraphics[width=0.48\linewidth]{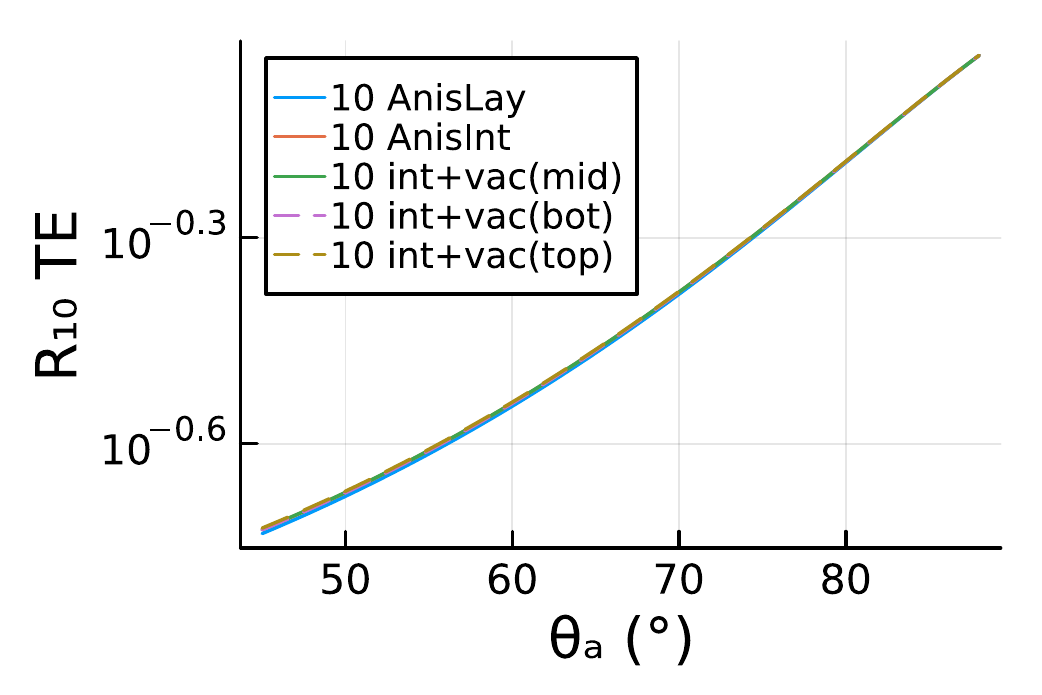}&
 \includegraphics[width=0.48\linewidth]{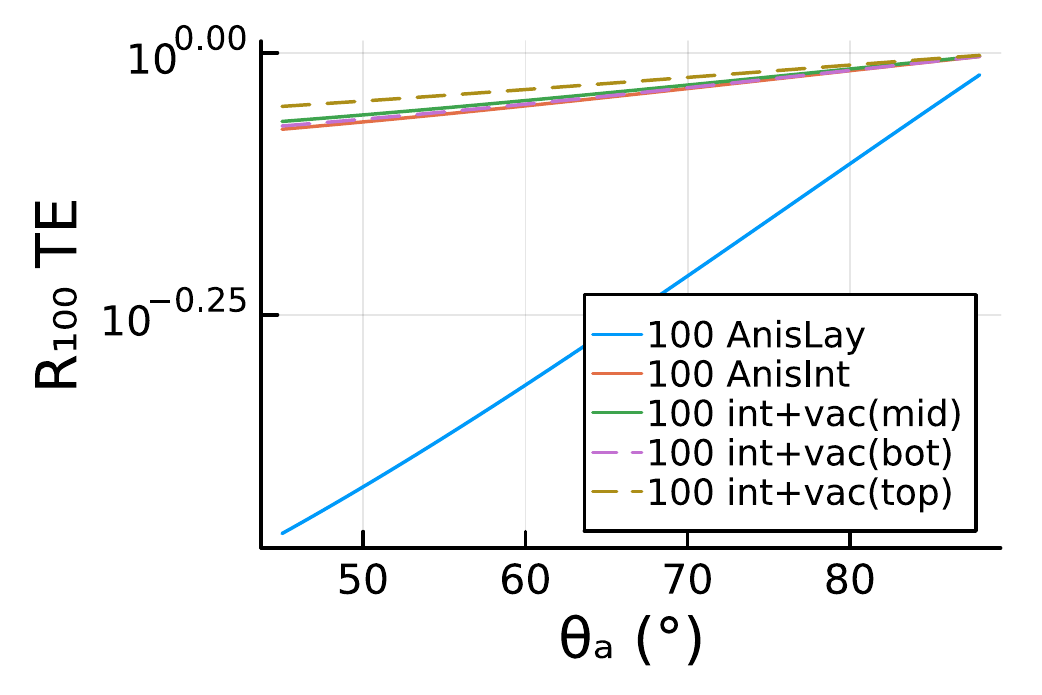}\\
 \includegraphics[width=0.48\linewidth]{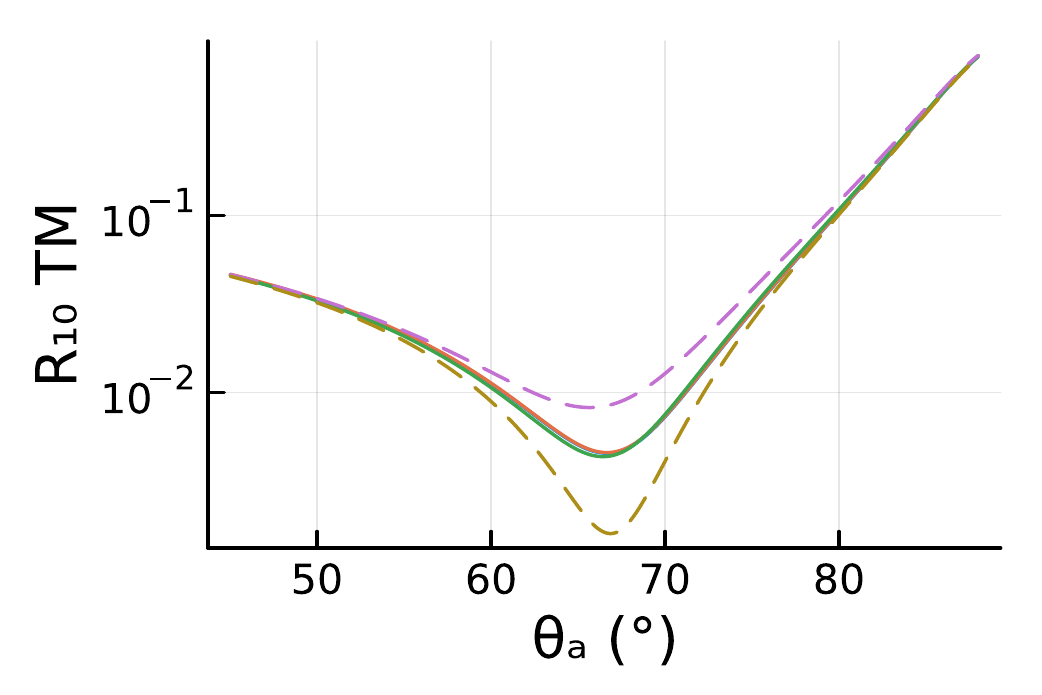}&
 \includegraphics[width=0.48\linewidth]{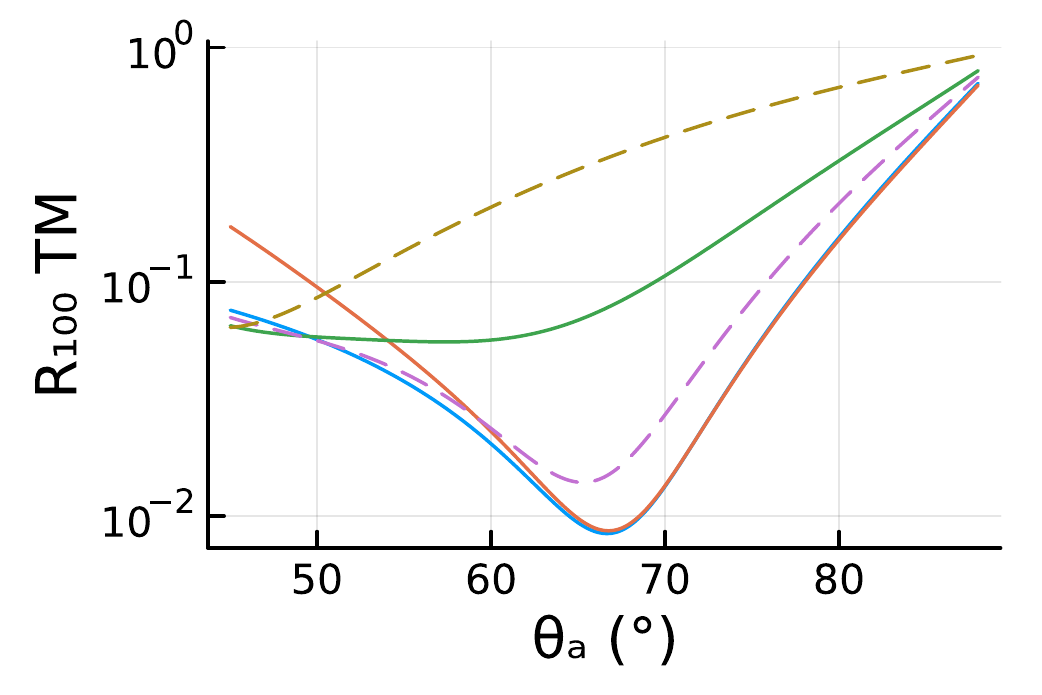}\\
\end{tabular}
\caption{Reflectance of multilayer systems of \chem{MoS_2} sheets immersed in a polymer with refractive index 1.4233. The five geometries of Fig.~\ref{Fig:multilayer} are considered. Left: 10 stacked layer. Right: 100 layers. Top, TE reflectance. Bottom, TM reflectance. }
\label{Fig:10L100LMoS2}
\end{figure}

Results are shown on Fig.~\ref{Fig:10L100L} and Fig.~\ref{Fig:10L100LMoS2}, respectively for graphene and \chem{MoS_2} sheets immersed in a polymer with refractive index \(n_a=1.4233\).
For 10 layers, the change of reflectance between the interface and the layer models is negligible with respect to the experimental precision when graphene is considered. Modelling with the configurations "vac+top/bot" provides strong differences around the Brewster angle already at 10 layers for \chem{MoS_2}.
For 100 layers, a change of the TM reflectance curve is clearly seen between the three configurations on the right of Fig.~\ref{Fig:multilayer} while the three models labeled \textit{interface}, \textit{layer} and \textit{int+vac(mid)} provide more similar curves in the case of graphene. This is not the case for \chem{MoS_2}, where  ``int+vac(mid)'' also differs from the \textit{interface} and \textit{layer} ones. The divergence between the {\ALM} and {\AIM} comes from their different angular dependency that is negligible for one layer but has an impact on 100 layers.
From these results we conclude that {\ALM} should be used for multilayer systems, as it is the
one that better reflects the physical distribution of dipolar response in the volume. Predictions made with the configuration ``int+vac(bot/top)'' should be avoided.

\section{Summary and conclusion}

We proposed a new model to describe the interaction between an electromagnetic wave and a bidimensional material or an heterostructure of stacked 2D layers, and we made the comparison with existing models describing the interaction with a single-layer. Our new layer model is based on intrinsic parameters. In particular,  the out-of -plane polarization response of the current sheet is related to \(D_z\) using the displacement susceptibility \(\xi_z\).
The {\ALM} leads to simple expressions of the transmission and reflection ratios that predict flux conservation in a losless medium. It offers a complete picture where complex phase jumps at the 2D interface can be deduced from the volume parameters, or conversely. This shows that a first order description at this level contains all the physics. The {\ALM} offers a dual vision of layers that can be continuous or correspond to a current sheet surrounded by vacuum.

Application of the {\ALM} to heterostructures was straightforward. It demonstrated that the {\ALM} cannot be replaced by a simplified model for a high number of layers in TM configuration.

The relations with the previous models allow to connect volume parameters to ab-initio calculation~\cite{semnani_nonlinear_2016} that can be performed in the 2D material.

Our approach shows that for an accurate modelling, the thickness of the layer is important. For a low number of layers, the in-plane dipole contributions can be summed. For the out-of-plane component, we have introduced the displacement susceptibility \(\xi\) allowing to assign surface values.

We believe that this work answers important questions for the community. In particular, we connect the different results published so far and provide the expressions to obtain intrinsic parameters from the different existing models. The use of these intrinsic parameters will be helpful to understand, characterize and tune properties of heterostructures.

\appendix
\section{Boundary conditions for TE and TM modes}
\label{Sec:BCTETM}

As detailed in~\cite{majerus_electrodynamics_2018}, boundary conditions for a TE (\(\TE\)) incident plane wave of frequency \(\omega/(2\pi)\), and wavevector \(\vec{k}=k_x\vec{1}_x+k_z\vec{1}_x\) on a 2D material located in plane \(z=0\), are (t stands for the transmitted fields, r for the reflected fields, and i for the incident ones)
\begin{eqnarray}
 E_y^t&=&E_y^i+E_y^r,\label{Eq:TE:t-r}\\
 k_z^tE_y^t&=&k_z^iE_y^i+k_z^rE_y^r+\irm\mu_0\omega^2\mathcal{P}_{y},
\label{Eq:TE:t+r}
\end{eqnarray}
while those for a TM (\(\TM\)) wave are
\begin{eqnarray}
 D_z^t&=&D_z^i+D_z^r-\irm{}k_x\mathcal{P}_x,\label{Eq:TM:t-r}\\
 \alpha_{it}^{\parallel}D_z^t&=&D_z^i-D_z^r+\irm{}k_x^2\frac{\varepsilon_x^i}{k_z^i\varepsilon_0}\mathcal{P}_{z}\label{Eq:TM:t+r},
\end{eqnarray}
with \(\vec{\mathcal{P}}\), the surface polarization field.

For symmetry reasons, it is most of the time possible to assume that the TE and the TM waves are not coupled by the surface polarization term, \textit{i.e.} \(\chi^s_{xz}=\chi^s_{yz}=\chi^s_{zx}=\chi^s_{zy}=0\).
In this case, Eqs.~(\ref{Eq:TE:t-r}) and (\ref{Eq:TM:t-r}); and  (\ref{Eq:TE:t+r}) and (\ref{Eq:TM:t+r}) write
\begin{eqnarray}
 t-r-1&=&\irm{}\varphi_{it},\label{A:Eq:BC:x}\\
 \alpha_{it}t+r-1&=&\irm\psi_{it}.
 \label{A:Eq:BC:yz}
\end{eqnarray}
Writing \(k_0^2=\omega^2/c^2\), with \(c\) and \(\varepsilon_0\), the speed of light and permittivity in vacuum,
\begin{eqnarray}
t_{\TE}=E_y^t/E_y^i,
\;
r_{\TE}=E_y^r/E_y^i,\label{Eq:tTErTE}\\
\alphaTE_{it} = \frac{k_{z,\TE}^{t}}{k_{z,\TE}^{i}},\;
k_{z,\TE}^2=\frac{\varepsilon_{x}}{\varepsilon_0}k_0^2-k_x^2,\label{Eq:kzTE}\\
\phiTE_{it}=0,
\;
\psiTE_{it}=\frac{k_0^2}{k_{z,\TE}^{i}} t_{\TE}\frac{\mathcal{P}_y}{\varepsilon_0\,E_y^{t}},\label{Eq:phipsiTE}\\
t_{\TM}=D_{z}^t/D_z^i,
\;
r_{\TM}=D_z^r/D_z^i,\label{Eq:tTMrTM}\\
\alphaTM_{it} = \frac{\varepsilon_{x}^{i}k_{z,\TM}^{t}}{\varepsilon_{x}^{t}k_{z,\TM}^{i}},
\;
k_{z,\TM}^2=\frac{\varepsilon_{x}}{\varepsilon_0}k_0^2-\frac{\varepsilon_{x}}{\varepsilon_z}k_x^2,\label{Eq:kzTM}\\
\phiTM_{it} 
      = \frac{k_{z,\TM}^{t}}{\varepsilon_x^t} t_{\TM}\frac{\mathcal{P}_x}{E_x^t},
\;
\psiTM_{it} 
      = \frac{k_x^2\varepsilon_x^i}{k_{z,\TM}^{i}\varepsilon_0} t_{\TM}
        \frac{\mathcal{P}_z}{D_z^t},\label{Eq:phipsiTM}.
\end{eqnarray}

\section{Energy conservation}

The layer matrix (\ref{Eq:LayerMatrix}) is conservative when \(k_z^m\) is a real quantity. However its first order expansion in \(\Phi\) has been criticized because it violates flux conservation for insulators~\cite{xu_optical_2021}.

While this is true, from an algebraic perspective, this is fortunately not the case if the calculation is limited to the first order.
To understand the origin of the problem, we consider (\ref{Eq:LmDetaileProd}) in the case where \(k_z^m\) is real. The interface matrices are not concerned with the Taylor expansion, and the flux violation can therefore not come from these matrices. The propagation matrix provides \(r=0\),  \(t=\exp(\irm{k_z^m}d_m)\) and \(\vert{t}\vert^2=1\). Flux conservation is therefore verified if \(k_z^m\) is real. However, when \(t\) is expanded to first order, we get \(t=1+\irm k_z^m d_m\) so that \(\vert{t}\vert^2= 1+ \left\vert{}k_z^m d_m\right\vert^2\) showing that the flux is no more conserved. The error is of second order, which is negligible for typical values of \(\Phi=k_z^md_m\), as shown in Sec.~\ref{Sec:discretemedium}, and makes sense from an analytical point of view.

Though this is not problematic, it would be more comfortable to work with an approximation that ensures flux conservation. One way to achieve this is to rewrite the imaginary exponential before applying Taylor expansion, to get
\begin{eqnarray}
\exp(\irm\Phi) &=& \frac{\exp(\irm\Phi/2)}{\exp(-\irm\Phi/2)}
= \frac{1+\irm\tan(\Phi/2)}
       {1-\irm\tan(\Phi/2)},\\
&=& \frac{1+\irm\Phi/2+\bigO{\Phi^3}}
       {1-\irm\Phi/2+\bigO{\Phi^3}}
\label{Eq:ConsApproxExp},
\end{eqnarray}
that is algebraically conservative when \(\Phi\) is real%
, as
\begin{equation}
\left\vert\frac{1+\irm\Phi/2}
       {1-\irm\Phi/2}\right\vert^2
=\frac{1+\Phi^2/4}
       {1+\Phi^2/4}=1
\label{Eq:ConsApproxPhi2}
.
\end{equation}

This higher order approximation is used in Sec.~\ref{Sec:layertoBC} to build the layer matrix, and in Sec.~\ref{Sec:fitting} to match the {\AIM} and the interface one.

\bibliographystyle{apsrev4-1}
\bibliography{2Dvs3D}

\makeatletter\@input{mksuppaux.tex}\makeatother

\end{document}